\documentclass[11pt]{article}
\usepackage[a4paper,hmargin=1.0in,vmargin=1.0in]{geometry}
\usepackage{mathtools,amsthm,amssymb,setspace,sectsty,bbm}
\usepackage[round]{natbib}
\usepackage[usenames,dvipsnames]{xcolor}
\usepackage[linktocpage=true,pagebackref=true,colorlinks,allcolors=blue,bookmarks=true,bookmarksopen,bookmarksnumbered]{hyperref}

\newtheoremstyle{DStheorem}
  {\topsep}
  {\topsep}
  {\itshape}
  {0pt}
  {\scshape}
  {.}
  { }
  {\thmname{#1}\thmnumber{ #2}\thmnote{ (#3)}}
\theoremstyle{DStheorem}
\newtheorem{theorem}{Theorem}[section]
\newtheorem{lemma}[theorem]{Lemma}
\newtheorem{claim}[theorem]{Claim}
\newtheorem{observation}[theorem]{Observation}

\let\oldproofname=\proofname
\renewcommand{\proofname}{\rm\sc{\oldproofname}}

\newcommand{\bs}[1]{\boldsymbol{#1}}
\newcommand{\bbR}{\mathbbm{R}}
\newcommand{\eps}{\epsilon}
\newcommand{\hmax}{H_{\max}}
\newcommand{\myadapt}{\mathcal{A}}
\newcommand{\mystat}{\mathcal{S}}
\newcommand{\mybound}{\mathcal{U}^*}
\newcommand{\mybern}{\mathrm{Bernoulli}}
\newcommand{\poly}{\mathrm{poly}}
\newcommand{\opt}{\mathrm{OPT}}
\newcommand{\ex}[1]{\mathbbm{E}\left[#1\right]}
\newcommand{\expar}[1]{\mathbbm{E}[#1]}

\newcommand{\exsubpar}[2]{\mathbbm{E}_{#1}[#2]}
\newcommand{\pr}[1]{\mathrm{Pr}\left[#1\right]}
\newcommand{\prpar}[1]{\mathrm{Pr}[#1]}

\newcommand{\geqst}{\succeq_{\mathrm{st}}}

\sectionfont{\large} \subsectionfont{\normalsize}
\allowdisplaybreaks
\makeindex

\newcounter{ggcounter}
\newcounter{dannycounter}

\begin{document}

\begin{titlepage}

\title{A Constructive Prophet Inequality Approach to \\
The Adaptive ProbeMax Problem}
\author{%
Guillermo Gallego\thanks{School of Data Science, The Chinese University of Hong Kong, Shenzhen, China, 518172. Email: {\tt gallegoguillermo@cuhk.edu.cn}. Supported by RGC project 16211619 and CRF project C6032-21G.}
\and
Danny Segev\thanks{Department of Statistics and Operations Research, School of Mathematical Sciences, Tel Aviv University, Tel Aviv 69978, Israel. Email: {\tt segevdanny@tauex.tau.ac.il}. Supported by Israel Science Foundation grant 1407/20.}}
\date{}
\maketitle

\setcounter{page}{200}
\thispagestyle{empty}

\begin{abstract}
In the adaptive ProbeMax problem, given a collection of  mutually-independent random variables $X_1, \ldots, X_n$, our goal is to design an adaptive probing policy for sequentially sampling at most $k$ of these variables, with the objective of maximizing the expected maximum value sampled. In spite of its stylized formulation, this setting captures numerous technical hurdles inherent to stochastic optimization, related to both information structure and efficient computation. For these reasons, adaptive ProbeMax has served as a test bed for a multitude of algorithmic methods, and concurrently as a popular teaching tool in courses and tutorials dedicated to recent trends in optimization under uncertainty.

The main contribution of this paper consists in proposing a novel method for upper-bounding the expected maximum reward of optimal adaptive probing policies, based on a simple min-max problem. Equipped with this method, we devise purely-combinatorial algorithms for deterministically computing feasible sets whose vicinity to the adaptive optimum is analyzed through prophet inequality ideas. Consequently, this approach allows us to establish improved constructive adaptivity gaps for the ProbeMax problem in its broadest form, where $X_1, \ldots, X_n$ are general random variables, making further advancements when $X_1, \ldots, X_n$ are continuous.
\end{abstract}


\end{titlepage}

\setcounter{page}{200}
\thispagestyle{empty}
\tableofcontents

\newpage
\setcounter{page}{1}
\section{Introduction}

In the last two decades, we have been witnessing a burst of theoretical advances surrounding stochastic combinatorial optimization, leading to innovative analytical methods and algorithmic techniques in a wide range of domains. While problems falling into this framework come in various forms and shapes, their common theme is that of optimizing in the presence of stochastic uncertainty, typically involving randomness in the model parameters, input structure, allowable actions, as well as in how these ingredients jointly interact. Due to the breadth and depth of this research arena, and due to its well-established connections to stochastic programming, Markov decision processes, and competitive analysis, we refer avid readers to selected books in this context \citep{HeymanS04, HentenryckR06, SchneiderK07, BirgeL11, ShapiroDR21, Powell22} and to the references therein for a deeper dive into these topics.

In this paper, we revisit one of the most eye-opening computational settings in the subfield of stochastic probing, commonly known as the adaptive ProbeMax problem. Indeed, in spite of its stylized formulation, this setting still captures numerous technical hurdles inherent to stochastic optimization, related to both information structure and efficient computation. For these reasons, adaptive ProbeMax has repeatedly been serving as an test bed for a multitude of algorithmic methods, which will be surveyed in Section~\ref{subsec:prev_work_open_questions}, and concurrently as a popular teaching tool in courses and tutorials dedicated to recent trends in optimization under uncertainty; see, e.g., \citep{Munagala16, Bansal16LN, Gupta18Tut, Kesselheim20LN, Hoefer21LN, Singla22LN}. In order to rigorously discuss existing work in this context, to highlight pending open questions, and to present our main contributions, we proceed by providing a complete mathematical description of the problem in question.

\subsection{Model formulation} \label{subsec:model_desc}

Let $X_1, \ldots, X_n$ be a collection of mutually-independent non-negative random variables with finite expectations $\mu_1, \ldots, \mu_n$. From an information-theoretical perspective, we assume that the distribution of each $X_i$ is known to the decision maker. That said, from a computational standpoint, our algorithms require two evaluation oracles, providing access to the cumulative distribution function $\prpar{ X_i \leq \cdot}$ and to conditional expectations of the form $\expar{ X_i | X_i \geq \cdot }$. As a side note, these assumptions have thoroughly been exploited, either explicitly or implicitly, in nearly all papers that will be mentioned later on.
\paragraph{Adaptive probing policies.} In a nutshell, our goal is to design an adaptive probing policy for sequentially sampling at most $k$ of the random variables $X_1, \ldots, X_n$, which will be referred to as rewards, with the objective of maximizing the expected maximum reward sampled. To formalize this setting, it is instructive to utilize dynamic programming notation. Specifically, let us consider a state description of the form $(\kappa,r,T)$, in which $\kappa \in [k]_0$ stands for the remaining number of rewards to be probed, $r \geq 0$ corresponds to the maximal value sampled thus far, and $T \subseteq [n]$ represents the collection of random variables that have not been probed yet. With this notation, an adaptive probing policy is simply a function $P : [k] \times \bbR_+ \times 2^{[n]} \to [n]$ that, given any state $(\kappa,r,T)$ with $\kappa \geq 1$, decides on the next reward $P(\kappa,r,T)$ to be probed out of the set of currently available rewards $T$.

\paragraph{System dynamics.} The random process we consider evolves along a sequence of discrete
stages, indexed by the remaining number $\kappa$ of rewards to be probed, in decreasing order. As such, transitions from one state to the next will be governed by the probing policy $P$ being examined as well as by the randomness in $X_1, \ldots, X_n$, according to the following dynamics:
\begin{itemize}
    \item {\em Initial state: $\kappa = k$.} At the beginning of stage $k$, each of the random variables $X_1, \ldots, X_n$ is available to be potentially probed, and we still have not collected any reward, meaning that our initial state is $(k,0,[n])$.

    \item {\em Probing step: $\kappa \geq 1$.} For each state $(\kappa,r,T)$ with $\kappa \geq 1$, the  policy $P$ picks one of the available rewards, $P(\kappa,r,T) \in T$, to be probed next. Once the reward $X_{ P(\kappa,r,T) }$ is sampled and its realization is revealed, we proceed to stage $\kappa - 1$ with the better reward out of $r$ and $X_{ P(\kappa,r,T) }$. It is important to emphasize that, since $X_1, \ldots, X_n$ are assumed to be independent, each of the yet-unprobed rewards $\{ X_i \}_{ i \in T \setminus \{ P(\kappa,r,T) \} }$ preserves its original distribution, regardless of how $X_{ P(\kappa,r,T) }$ is realized. Therefore, the probing policy $P$ maps our current state to a random state, given by
    \begin{equation} \label{eqn:transition_rule}
    (\kappa,r,T) ~~\xmapsto{~~P~~}~~ (\kappa-1, \max \{r, X_{ P(\kappa,r,T) } \}, T \setminus \{ P(\kappa,r,T) \}) \ .
    \end{equation}

    \item {\em Terminal state: $\kappa = 0$.} By the preceding discussion, states of the form $(0,r,T)$ will be reached as soon as $k$ rewards are sampled in total. In this case, our transition rule straightforwardly ensures that $r$ represents the maximum value sampled along the way.
\end{itemize}

\paragraph{Objective function.} To conveniently write the expected maximum value sampled by any given policy, it is useful to work with recursive expressions, where we make use of $\myadapt(\kappa,r,T)$ to designate the expected maximum reward attained by the policy $P$, starting at state $(\kappa,r,T)$. Here, ``$\myadapt$'' is meant to emphasize the adaptive nature of such policies, differentiating them from their static counterparts that will be introduced later on. In view of the transition rule~\eqref{eqn:transition_rule} for the general case of $\kappa \geq 1$, the latter function can be recursively written as
\[ \myadapt_P(\kappa,r,T) ~~=~~ \ex{ \myadapt_P(\kappa-1, \max \{r, X_{ P(\kappa,r,T) } \}, T \setminus \{ P(\kappa,r,T) \}) } \ , \]
where the expectation above is taken over the randomness in $X_{ P(\kappa,r,T) }$. In the terminal case of $\kappa = 0$, we clearly have $\myadapt_P(0,r,T) = r$.

With these definitions, in the adaptive ProbeMax problem, we wish to compute an adaptive probing policy $P$ whose expected maximum reward $\myadapt(P)$ is maximized. The latter measure stands for our expected maximum reward with respect to the initial system state, prior to probing any of the random variables $X_1, \ldots, X_n$, meaning that $\myadapt(P) = \myadapt_P(k,0,[n])$. In the sequel, $P^*$ will denote an arbitrary optimal adaptive policy, with $\myadapt^* = \myadapt(P^*)$ being its expected maximum reward.

\paragraph{Static ProbeMax.} Moving forward, it is instructive to briefly discuss the so-called static formulation of this setting. Here, our objective is to compute a subset of $k$ random variables, aiming to maximize the expected maximum reward of this subset. To formalize this notion,  let ${\cal K} = \{S \subseteq [n]: |S| =   k\}$ be the family of subsets with cardinality exactly $k$, and for every $S \subseteq [n]$, let $M(S) = \max_{i \in S} X_i$ be its random maximum value. Then, the static ProbeMax problem asks to identify a subset $S \in {\cal K}$ for which $\expar{ M(S) }$ is maximized. Analogously to the adaptive version, we make use of $\mystat^* = \max_{S \in {\cal K}} \expar{ M(S) }$ to denote the optimum value of a given instance in the static case.

\subsection{Existing work and open questions} \label{subsec:prev_work_open_questions}

The vast majority of algorithmic work around the adaptive ProbeMax problem has focused on the design of constant-factor approximations through non-adaptive policies, establishing a sequence of improved adaptivity gaps in this context. The latter term refers to the worst-possible ratio between the adaptive optimum $\myadapt^*_I$ and the static one $\mystat^*_I$, over all problem instances $I$, namely, $\sup_I \{ \frac{ \myadapt^*_I }{ \mystat^*_I } \}$. In what follows, we discuss the main technical approaches taken in order to attain these results, shedding some light on their scope, advantages, and downsides.

\paragraph{Approach 1: LP-based methods.} \cite{GuhaM07} demonstrated the surprising power of static policies for stochastic probing under general packing constraints. Specifically, their approach considers the case where $X_1, \ldots, X_n$ are discrete random variables with finite support, say ${\cal V}$. In this case, an upper bound on the adaptive optimum $\myadapt^*$ was shown to be attainable by solving the following linear relaxation:
\[ \begin{array}{llll}
(\mathrm{LP}) \qquad \qquad & \max & {\displaystyle \sum_{i \in [n]} \sum_{r \in {\cal V}} r \cdot y_{iv}} \\
& \text{s.t.} & {\displaystyle \sum_{i \in [n]} \sum_{r \in {\cal V}} y_{iv} \leq 1} \\
& & {\displaystyle \sum_{i \in [n]} x_i = k} \\
&&  y_{iv} \leq \prpar{X_i = v } \cdot x_i \qquad \qquad  & \forall \, i \in [n], \, r \in {\cal V} \\
& & x_i, y_{iv} \in [0,1] \qquad & \forall \, i \in [n], \, r \in {\cal V}
\end{array} \]
Given an optimal fractional solution, \citeauthor{GuhaM07} proposed a randomized rounding procedure for defining a distribution ${\cal D}$ over feasible subsets, such that $\exsubpar{S \sim {\cal D}}{ M(S) } \geq \frac{  \opt(\mathrm{LP}) }{ 8 } \geq \frac{ \myadapt^* }{ 8 }$. This result directly translates to an adaptivity gap of at most $8$. Subsequently, the work of \cite{GuptaN13} showcased the usefulness of contention resolution schemes \citep{ChekuriVZ14} for stochastic probing. In regard to adaptive ProbeMax, their approach leads to an improved rounding procedure for creating a distribution ${\cal D}$ over feasible subsets with $\exsubpar{S \sim {\cal D}}{ M(S) } \geq \frac{  \opt(\mathrm{LP}) }{ 3 }$, implying an adaptivity gap of at most $3$. Yet another LP-based method is that of \cite{GuhaMS10}, who devised a Lagrangian relaxation approach, through which an adaptivity gap of $3$ can be attained. In contrast to earlier ideas in this context, their algorithmic method applies when $X_1, \ldots, X_n$ are general random variables, assuming oracle access to certain distributional properties, and does not require solving any linear program or employing randomization. Instead, by exploiting very simple queries, \citeauthor{GuhaMS10} showed how to deterministically construct a set $S \in {\cal K}$, ending up with an expected maximum reward of $\expar{ M(S) } \geq \frac{ \opt(\mathrm{LP}) }{ 3+\eps }$.

From this point on, we say that a given adaptivity gap is constructive when it is accompanied by a polynomial-time algorithm, either deterministic or randomized, for explicitly specifying a matching  non-adaptive probing policy, possibly up to $1 + \eps$. In this regard, all adaptivity gaps mentioned thus far are constructive in nature. Moreover, to our knowledge, \cite{GuhaMS10} still hold the currently best constructive gap for arbitrary random variables, achievable via any method.

\paragraph{Approach 2: Submodularity-based methods.} Undoubtedly, one of the most powerful machineries for tackling adaptive probing problems is that of maximizing stochastic submodular functions subject to matroid constraints, as studied by \cite{AsadpourN16}. From this perspective, the important observation is that $F(S) = \expar{ M(S) } = \expar{  \max_{i \in S} X_i }$ is a monotone submodular set function. For any such function, its multilinear extension $\bar{F} : [0,1]^n \to \bbR$ is given by
\[ \bar{F}( x ) ~~=~~ \sum_{S \subseteq [n]} \left( \prod_{i \in S} x_i \right) \cdot \left( \prod_{i \notin S} (1 - x_i) \right) \cdot F(S) \ . \]
Interestingly, the pipage rounding method of \cite{CalinescuCPV11} can be employed to compute, for any vector $x \in [0,1]^n$ with $\| x \|_1 \leq k$, a corresponding set $S_x \in {\cal K}$ such that $F(S_x) \geq \bar{F}(x)$. While we state this finding for $k$-uniform matroids, it actually applies to arbitrary matroids. A fundamental result due to \citeauthor{AsadpourN16} resides in proving that, when the random variables $X_1, \ldots, X_n$ are absolutely continuous, there exists a vector $x$ for which $\expar{ M(S_x) } = F(S_x) \geq (1 - \frac{ 1 }{ e }) \cdot \myadapt^*$. In turn, we obtain an upper bound of $\frac{ e }{ e-1 } \approx 1.58$ on ProbeMax's adaptivity gap, which stands as the currently best known gap, although it is generally not constructive, as explained below. Still, it is worth pointing out that this result actually applies to general random variables, since each $X_i$ can be substituted by $\tilde{X}_i = X_i + \delta_i$, where $\delta_i$ is some ``tiny'' noise, say $\delta_i \sim U(0,\Delta)$ with $\Delta \ll \max_{i \in [n]} \mu_i$. This way, at the expense of introducing negligible errors in the expected maximum reward of any probing policy (adaptive or non-adaptive), one ensures that  $\tilde{X}_i$ is absolutely continuous and non-negative.

Noting that this adaptivity gap is existential in nature, \citeauthor{AsadpourN16} proposed the stochastic continuous greedy algorithm, guaranteed to identify a subset $S \in {\cal K}$ for which $\expar{ M(S) } \geq (1 - \frac{ 1 }{ e } - \eps) \cdot \myadapt^*$ with high probability. However, this approach admits a polynomial-time implementation only subject to additional technical assumptions (Lipschitz continuity, bounded variance), which are required in order to design a sampling-based oracle for the multilinear extension $\bar{F}$. As an alternative, one can resort to the polynomial-time approximation schemes of \cite{ChenHLLLL16} and \cite{Segev021} for static ProbeMax. These approaches would deterministically construct a set $S \in {\cal K}$ with an expected maximum reward of $\expar{ M(S) } \geq (1 - \eps) \cdot \mystat^* \geq (1 - \frac{ 1 }{ e } - \eps) \cdot \myadapt^*$, albeit at the expense of further assuming that $X_1, \ldots, X_n$ are bounded, due to their Bernoulli-decomposition-based discretization method.

\paragraph{Approach 3: Direct decision-tree arguments.} While still discussing broad-spectrum machineries, it is important to bring up the work of \cite{GuptaNS17} and \cite{Bradac0Z19} on maximizing stochastic submodular functions subject to prefix-closed probing constraints. Stated in terms of our particular setting, when $X_1, \ldots, X_n$ are discrete random variables with finite support, \cite{GuptaNS17} proved that taking an appropriately-chosen random path down the optimal decision tree forms a randomized non-adaptive policy whose expected maximum reward is within factor $3$ of the adaptive optimum. The latter finding was sharpened by \cite{Bradac0Z19}, who obtained an improved adaptivity gap of $2$. Once again, we mention that both adaptivity gaps are existential, since the optimal decision tree is clearly unknown.

\paragraph{Approach 4: Block-adaptive policies.} On a different front, unrelated to adaptivity gaps, it is imperative to mention the recent breakthrough of \cite{FuLX18}. Here, under the assumptions that the random variables $X_1, \ldots, X_n$ are bounded and that we have oracle access to certain distributional properties, the authors established that the family of so-called block-adaptive policies approximate the expected maximum reward $\myadapt^*$ of an optimal adaptive policy within factor $1-\eps$. Moreover, \citeauthor{FuLX18} proposed an $O( n^{ 2^{ \poly(1/\eps) } }  )$-time algorithm for computing an optimal block-adaptive policy by means of dynamic programming. Subsequently, \cite{Segev021} improved the latter running time to $O( 2^{ 2^{ \poly(1/\eps) } } n^{ O(1) } )$, through rounding appropriate LP-relaxations of the multi-dimensional Santa Claus problem. That said, due to their double-exponential dependency on the accuracy level $\eps$, these approaches are mostly theoretical.

\paragraph{Motivating questions.} In light of the preceding discussion, the primary open questions that motivate our work aim to fill several interrelated voids in the current literature, along the following axes:
\begin{itemize}
    \item {\em Upper bounds.} Beyond linear relaxations, multilinear extensions, and decision-tree arguments, are there alternative methods for efficiently obtaining tight upper bounds on the adaptive optimum?

    \item {\em Computational efficiency and randomization.} Can we leverage such upper bounds in order to efficiently identify feasible sets that well-approximate the best adaptive policy? Can we come up with deterministic constructions, or perhaps randomization is a true necessity?

    \item {\em Generality.} Can we establish improved constructive adaptivity gaps for general random variables? What about continuous ones? To our knowledge, the currently best adaptivity gap in both contexts is still $3$, via the Lagrangian relaxation approach of \cite{GuhaMS10}.
\end{itemize}

\subsection{Contributions and techniques} \label{subsec:contributions}

The main contributions of this paper consist in proposing a novel method for upper-bounding the expected maximum reward of optimal adaptive probing policies, based on a simple min-max problem. Equipped with this method, we devise purely-combinatorial algorithms for deterministically computing feasible sets whose vicinity to the adaptive optimum is analyzed through prophet inequality ideas. Consequently, this approach allows us to establish improved constructive adaptivity gaps for the ProbeMax problem in its broadest form, where $X_1, \ldots, X_n$ are general random variables, making further advancements when $X_1, \ldots, X_n$ are continuous. In what follows, we present a high-level account of our main results, touching upon selected technical ideas along the way.

\paragraph{The min-max upper bound.} To better understand the simplicity of our construction, consider some feasible set, $S \in {\cal K}$. Clearly, for any $r \in \bbR$, a straightforward upper bound on the maximum reward $M(S) = \max_{i \in S} X_i$ is given by
\[ M(S) ~~\leq~~ r + [M(S) - r]^+ ~~\leq~~ r + \sum_{i \in S} [X_i - r]^+ \ . \]
Letting $H(r,S) = r + \sum_{i \in S} \expar{ [X_i - r]^+ }$ be the expected value of the right-hand-side, focusing on the static optimum, we know that $\mystat^* = \max_{S \in {\cal K}} \expar{ M(S) } \leq \max_{S \in {\cal K}} H(r,S)$. Hence, the best-possible bound of this form on the static optimum is derived by minimizing the latter expression over $r$, thereby obtaining
\begin{equation} \label{eqn:definition_U} \tag{MinMax}
\mybound ~~=~~ \min_{r \in \bbR} \max_{S \in {\cal K}} H(r,S) \ .
\end{equation}
Quite surprisingly, in Section~\ref{sec:single_var_bound}, we exploit dynamic programming based characterizations of optimal adaptive policies to prove that $\mybound$ actually constitutes an upper bound on the adaptive optimum $\myadapt^*$, as formally stated in Theorem~\ref{thm:rel_dk_uk} below. Moreover, we establish a number of basic properties regarding some of the functions appearing in problem~\eqref{eqn:definition_U}, mostly related to convexity and differentiability, which will be useful for computational and analytical purposes.

\begin{theorem} \label{thm:rel_dk_uk}
$\myadapt^* \leq \mybound$.
\end{theorem}

\paragraph{General random variables: Adaptivity gap of $2$.} In Section~\ref{sec:adaptivity_2}, we examine the broadest possible setting, in which $X_1, \ldots, X_n$ are general random variables. Here, letting $r^*$ be an optimal solution to problem~\eqref{eqn:definition_U}, we prove that there exists a corresponding feasible set $\tilde{S} \in {\cal K}$, optimal with respect to the inner maximization problem $\max_{S \in {\cal K}} H(r^*,S)$, for which the following adversarial-order prophet-inequality-type result holds:
\begin{quote}
    {\em By inspecting the random variables $\{ X_i \}_{i \in \tilde{S}}$ in arbitrary order, and employing a root-threshold-based stopping policy $T_{ \tilde{S} }$, our expected reward is $\expar{ X_{ T_{ \tilde{S} } } } \geq \frac{ \mybound }{ 2 }$.}
\end{quote}
As an immediate corollary, since $\expar{ M(\tilde{S}) } \geq \expar{ X_{ T_{ \tilde{S} } } }$, we establish an adaptivity gap of at most $2$ for the most general formulation of the adaptive ProbeMax problem.

\begin{theorem} \label{thm:main_theorem_general}
When $X_1, \ldots, X_n$ are general random variables, $\frac{ \myadapt^* }{ \mystat^* } \leq 2$.
\end{theorem}

At a high level, these results are derived through structural arguments related to the upper envelope function $r \mapsto \max_{S \in {\cal K}} H(r,S)$, along with a deep dive into the type of guarantees that can be extracted from the root-threshold-based stopping policy in this context. While our analysis is self-contained, readers may benefit from consulting relevant surveys on prophet inequalities, such as those of \cite{HillK92survey}, \cite{Lucier17}, and \cite{CorreaFHOV18}. 

\paragraph{Continuous random variables: Adaptivity gap of ${\frac{ e }{ e-1 } \approx 1.58}$.} In Section~\ref{sec:adapt_gap_e_em1}, we study the more lenient scenario, where $X_1, \ldots, X_n$ are assumed to be continuous random variables, in the sense of having cumulative distribution functions that are continuous everywhere. Somewhat informally, letting $r^*$ be an optimal solution to problem~\eqref{eqn:definition_U}, we prove
that a linear extension of its corresponding inner problem, where one maximizes $\bar{H}(r^*,\psi) = r^* + \sum_{i \in S} \expar{ [X_i - r^*]^+ } \cdot \psi_i$ over $\Psi = \{ \psi \in [0,1]^n : \| \psi \|_1 = k \}$, admits an almost-integer optimal solution $\psi^*$ with a very specific derivative structure. Guided by this solution, we show that there exists a feasible set $\tilde{S} \in {\cal K}$ satisfying the next free-order prophet-inequality-type property:
\begin{quote}
    {\em By inspecting the random variables $\{ X_i \}_{i \in \tilde{S}}$ in order of weakly-decreasing $\expar{ X_i | X_i \geq r^* }$, and employing the stopping policy $T_{ \tilde{S} }$ where $r^*$ serves as a threshold, our expected reward is $\expar{ X_{ T_{ \tilde{S} } } } \geq (1 - \frac{ 1 }{ e }) \cdot \mybound$.}
\end{quote}
Once again, since $\expar{ M(\tilde{S}) } \geq \expar{ X_{ T_{ \tilde{S} } } }$, we attain an improved adaptivity gap of at most $\frac{ e }{ e-1 }$ in this setting.
\begin{theorem} \label{thm:main_theorem_cont}
When $X_1, \ldots, X_n$ are continuous random variables, $\frac{ \myadapt^* }{ \mystat^* } \leq \frac{ e }{ e-1 }$.
\end{theorem}

Deferring the finer details of these results to be discussed in Section~\ref{sec:adapt_gap_e_em1}, it is still worth mentioning that our analysis is operating on two fronts. The first of these directions introduces and studies the linear extension $\max_{\psi \in \Psi} \bar{H}(r^*, \psi)$, ending up with an explicit construction of the structured optimal solution $\psi^*$. Concurrently, the second direction develops the necessary theory behind our new free-order prophet inequality, which may very well be applicable in additional settings.

\paragraph{Algorithmic considerations.} By now, the keen-eyed reader must have noticed that the current statements of Theorems~\ref{thm:main_theorem_general} and~\ref{thm:main_theorem_cont} do not improve on the best known constructive adaptivity gap of $3$, due to \cite{GuhaMS10}, since they are still not algorithmic in nature. Indeed, when $X_1, \ldots, X_n$ are general random variables, we merely claim that the set $\tilde{S}$ forms an optimal solution to $\max_{S \in {\cal K}} H(r^*,S)$; however, the latter problem may have exponentially-many such solutions. Similarly, when $X_1, \ldots, X_n$ are continuous, we are claiming that the linear extension $\max_{\psi \in \Psi} \bar{H}(r^*, \psi)$ admits an optimal solution $\psi^*$ of very specific structure; here, the collection of such solutions may not even be countable. For these reasons, in Sections~\ref{sec:adaptivity_2} and~\ref{sec:adapt_gap_e_em1}, we further explain how to convert our analysis to purely-combinatorial polynomial-time algorithms, explicitly constructing feasible sets whose expected maximum reward matches the above-mentioned adaptivity gaps up to a factor of $1 + \eps$. 
\section{The Min-Max Bound and its Properties} \label{sec:single_var_bound}

In this section, we present a dynamic programming characterization of optimal adaptive policies, allowing us to prove that the min-max expression $\mybound$ indeed forms an upper bound on the adaptive optimum $\myadapt^*$, as stated in Theorem~\ref{thm:rel_dk_uk}. Subsequently, we further examine problem~\eqref{eqn:definition_U} and derive several convexity and differentiability results that will be useful in efficiently computing the upper bound $\mybound$, in presenting our algorithmic ideas, and in analyzing their performance guarantees.

\subsection{Proof of Theorem~\ref{thm:rel_dk_uk}} \label{subsec:proof_thm_rel_dk_uk}

\paragraph{Optimality via dynamic programming.} To show that $\mybound = \min_{r \in \bbR} \max_{S \in {\cal K}} H(r,S)$ is an upper bound on the adaptive optimum, we begin by explaining how an optimal adaptive policy $P^*$ generally operates. To this end, according to the system dynamics described in Section~\ref{subsec:model_desc}, the expected maximum reward attained by this policy at any state $(\kappa,r,T)$ can be recursively written as
\[ \myadapt_{P^*}(\kappa,r,T) ~~=~~ \ex{ \myadapt_{P^*}(\kappa-1, \max \{r, X_{ P^*(\kappa,r,T) } \}, T \setminus \{ P^*(\kappa,r,T) \}) } \ .  \]
Therefore, the optimal policy clearly picks the next reward $P^*(\kappa,r,T) \in T$ to be probed as the one that maximizes the latter expectation, arbitrarily breaking ties. In other words,
\begin{equation} \label{eqn:DP_optimal_policy}
\myadapt_{P^*}(\kappa,r,T) ~~=~~ \max_{i \in T} \ex{ \myadapt_{P^*}(\kappa-1, \max \{r, X_i \}, T \setminus \{ i \}) } \ , 
\end{equation}
with the convention that $\myadapt_{P^*}(0,r,T) = r$ for every terminal state $(0,r,T)$.

\paragraph{The recursive claim.} We proceed by inductively proving that, for every $\kappa \in [k]_0$, $r \in \bbR$, and $T \subseteq [n]$, one has
\begin{equation} \label{eqn:main_rel_dk_uk}
\myadapt_{P^*}(\kappa,r,T) ~~\leq~~ \min_{\rho \in \bbR} \left\{ [ r - \rho ]^+ + \max_{S \subseteq T : |S| = \kappa} H(\rho, S) \right\} \ .
\end{equation}
This bound allows us to argue that
$\myadapt^* \leq \mybound$, since
\begin{eqnarray*}
\myadapt^* & = & \myadapt_{P^*}(k,0,[n]) \\
& \leq &\min_{\rho \in \bbR} \left\{ [ 0 - \rho ]^+ + \max_{S \subseteq [n] : |S| = k} H(\rho, S) \right\} \\
& = &\min_{\rho \in \bbR} \left\{ [ - \rho ]^+ + \max_{S \in {\cal K}} H(\rho, S) \right\} \\
& = & \min_{\rho \in \bbR} \max_{S \in {\cal K}} H(\rho, S) \\
& = & \mybound \ .
\end{eqnarray*}
Here, the inequality above is obtained by instantiating~\eqref{eqn:main_rel_dk_uk} with $\kappa = k$, $r = 0$, and $T = [n]$. In addition, the third equality holds since the function $\rho \mapsto [ - \rho ]^+ + \max_{S \in {\cal K}} H(\rho, S)$ does not have negative minimizers. Indeed, by recalling that $H(\rho,S) = \rho + \sum_{i \in S} \expar{ [X_i - \rho]^+ }$, it is easy to verify that $[ - \rho ]^+ + \max_{S \in {\cal K}} H(\rho, S) > H(0, S)$ for every $\rho < 0$.

\paragraph{Base case: $\bs{\kappa = 0}$.} In this case, for every $r \in \bbR$ and $T \subseteq [n]$, we have
\[ \min_{\rho \in \bbR} \left\{ [ r - \rho ]^+ + \max_{S \subseteq T : |S| = 0}  H(\rho, S) \right\} ~~=~~ \min_{\rho \in \bbR} \left\{ [ r - \rho ]^+ + \rho \right\} ~~\geq~~ r ~~=~~ \myadapt_{P^*}(0,r,T) \ . \]

\paragraph{General case: $\bs{\kappa \geq 1}$} Here, for every $r \in \bbR$, $\rho \in \bbR$, and $T \subseteq [n]$, we have by equation~\eqref{eqn:DP_optimal_policy},
\begin{eqnarray*}
\myadapt_{P^*}(\kappa,r,T) & = & \max_{i \in T} \ex{ \myadapt_{P^*}(\kappa-1, \max \{r, X_i \}, T \setminus \{ i \}) } \\
& \leq &\max_{i \in T} \ex{ [ \max \{ r, X_i \} - \rho ]^+ + \max_{S \subseteq T \setminus \{i\} : |S| = \kappa-1}  H(\rho, S) } \\
& \leq &\max_{i \in T} \ex{ [ r - \rho ]^+ + [ X_i - \rho ]^+ + \max_{S \subseteq T \setminus \{i\} : |S| = \kappa-1}   H(\rho, S)  } \\
& = & [ r - \rho ]^+ + \max_{i \in T} \left\{  \ex{ [ X_i - \rho ]^+ } + \max_{S \subseteq T \setminus \{i\} : |S| = \kappa-1} H(\rho, S) \right\} \\
& = & [ r - \rho ]^+ + \max_{i \in T}  \max_{S \subseteq T \setminus \{i\} : |S| = \kappa-1} H(\rho, S \cup \{ i \}) \\
& = & [ r - \rho ]^+ + \max_{S \subseteq T : |S| = \kappa} H(\rho, S ) \ ,
\end{eqnarray*}
where the first inequality follows from the induction hypothesis~\eqref{eqn:main_rel_dk_uk}. Now, since this bound applied to every $\rho \in \bbR$, we indeed get \[ \myadapt_{P^*}(\kappa,r,T) ~~\leq~~ \min_{\rho \in \bbR} \left\{ [ r - \rho ]^+ + \max_{S \subseteq T : |S| = \kappa} H(\rho, S) \right\} \ . \]

\subsection{Properties of \texorpdfstring{$\bs{H(\cdot,S)}$}{} and \texorpdfstring{$\bs{\hmax}$}{}} \label{subsec:properties_HrS}

\paragraph{Convexity.} In order to simplify some upcoming notation, for $r \in \bbR$ and $S \subseteq [n]$, let us define the function $G(r,S) = \sum_{i \in S} G_i(r)$, where $G_i(r) = \expar{ [X_i - r]^+ }$. We first observe that, for every $i \in [n]$, the function $r \mapsto [X_i-r]^+$ is convex over $\bbR$. In turn, since convexity is preserved by expectations, it follows that $G_i(r)$ is convex as well. Finally, recalling that $H(r,S) = r + \sum_{i \in S} \expar{ [X_i - r]^+ } = r + \sum_{i \in S} G_i(r)$, we conclude that $H(\cdot,S)$ is convex, as a summation of such functions.

\begin{observation} \label{obs:prop_H2_min}
For any $S \in {\cal K}$, the function $H(\cdot,S)$ is convex over $\bbR$. 
\end{observation}

\paragraph{Approximately solving problem~\eqref{eqn:definition_U}.}  Recalling that $\mybound = \min_{r \in \bbR} \max_{S \in {\cal K}} H(r,S)$, let us separately consider its inner maximization problem, which will be designated by $\hmax( r ) = \max_{S \in {\cal K}} H(r,S)$. In addition, $R_{\hmax}$ will stand for its set of minimizers, meaning that
\[ R_{\hmax} ~~=~~ \left\{ r \in \bbR : \hmax(r) \leq \hmax(\hat{r}) \, \, \forall \hat{r} \in \bbR \right\} \ . \]
The next claim, whose proof is provided in Section~\ref{subsec:proof_lem_prop_calH_min}, shows that $\hmax(\cdot)$ is convex and identifies a closed interval over which its minimum value is attained. Below, $\mu_{\max} = \max_{i \in [n]} \mu_i$ stands for the maximum expectation of the random variables $X_1, \ldots, X_n$.

\begin{lemma} \label{lem:prop_calH_min}
$\hmax(\cdot)$ is convex over $\bbR$, with $R_{\hmax} \cap [0, n \mu_{\max}] \neq \emptyset$.
\end{lemma}

Given this result, assuming one can indeed evaluate $\hmax(\cdot)$, we can efficiently identify an approximate minimizer with respect to problem~\eqref{eqn:definition_U} by employing any of the countless methods for optimizing single-variable convex functions over a closed interval. To better understand this statement, we separately discuss these two points:
\begin{itemize}
    \item {\em Evaluation.} Noting that $\hmax(r) = r + \max_{S \in {\cal K}} \sum_{i \in S} G_i(r)$, maximizers of the latter term are clearly subsets corresponding to the $k$ largest values out of $\{ G_i(r) \}_{ i \in [n]}$, breaking ties arbitrarily. Therefore, having access to the cumulative distribution functions $\prpar{ X_i \leq \cdot}$ and to conditional expectations of the form $\expar{ X_i | X_i \geq \cdot }$, as assumed in Section~\ref{subsec:model_desc}, we know each of the values 
    \[ G_i(r) ~~=~~ \ex{ [X_i - r]^+ } ~~=~~ \pr{ X_i \geq r } \cdot (\ex{ X_i | X_i \geq r} - r) \ , \]
    implying that $\hmax(\cdot)$ can be efficiently evaluated.
    
    \item {\em Optimization.} Given Lemma~\ref{lem:prop_calH_min}, classical methods such as Golden-section search allow us to determine a subinterval $[r^-, r^+] \subseteq [0, n \mu_{\max}]$ that contains at least one minimizer $r^* \in R_{\hmax}$; to this end, $O( \log ( \frac{ n \mu_{\max} }{ r^+ - r^- } ) )$ queries of the function $\hmax(\cdot)$ will be required. Practically speaking, in Section~\ref{sec:adaptivity_2}, where we consider general random variables, a gap of $r^+ - r^- = O( \frac{ \eps \mu_{\max} }{ k } )$ suffices for our purposes, meaning that only $O( \log ( \frac{ n k }{ \eps } ) )$ such queries are needed. In Section~\ref{sec:adapt_gap_e_em1}, where we study continuous random variables, a similar gap suffices as well. However, in this case, to avoid complicating an already-involved analysis, we will work directly with an arbitrary minimizer $r^*$.
\end{itemize}

\paragraph{Differentiability of  $\bs{H(\cdot,S)}$ in the continuous case.} While Observation~\ref{obs:prop_H2_min} shows that each of the functions $H(\cdot,S)$ is convex, elementary examples demonstrate that they may not be differentiable. Still, when $X_1, \ldots, X_n$ are continuous random variables, we argue that $H(\cdot,S)$ is in fact differentiable, admitting a very specific derivative form that will be useful later on. This notion is formalized in the next claim, whose proof appears in Section~\ref{subsec:proof_lem_derivative_H}.

\begin{lemma} \label{lem:derivative_H}
Suppose that the random variables $X_1, \ldots, X_n$ are continuous. Then, for any  subset $S \subseteq [n]$, the function $H(\cdot,S)$ is differentiable, with $\frac{ d }{ dr } H(r,S) = 1 - \sum_{i \in S} \pr{ X_i \geq r }$.
\end{lemma}

\subsection{Proof of Lemma~\ref{lem:prop_calH_min}} \label{subsec:proof_lem_prop_calH_min}

By Observation~\ref{obs:prop_H2_min}, the function $H(\cdot,S)$ is convex over $\bbR$, for every subset $S \in {\cal K}$. Therefore, since $\hmax( r ) = \max_{S \in {\cal K}} H(r,S)$, it follows that $\hmax$ is convex  as well. Now, regarding its set of minimizers $R_{\hmax}$, we argue that
\begin{equation} \label{eqn:hmax_min_sup}
\min_{r \in [0,n \mu_{\max}]} \hmax(r) ~~\leq~~ \inf_{r \in \bbR} \hmax(r) \ .
\end{equation}
As a side note, taking the minimum on the left-hand-side is well-defined, since we have already shown that $\hmax$ is convex over $\bbR$; since any convex function is continuous over the interior of its domain, it follows that $\min_{r \in [0,n \mu_{\max}]} \hmax(r)$ is indeed attained. To verify inequality~\eqref{eqn:hmax_min_sup}, note that $\hmax(0) \leq \sum_{i \in [n]} \mu_i \leq m \mu_{\max}$. Therefore, when $r > n \mu_{\max}$, we have for example $H(r,[k]) \geq r > n \mu_{\max}$, implying that $\hmax(r) >n \mu_{\max} \geq \hmax(0)$. On the opposite end, when $r < 0$, we have $H(r,S) = -(k-1) \cdot r + \sum_{i \in S} \expar{ X_i } \geq H(0,S)$ for every $S \in {\cal K}$, implying that $\hmax(r) \geq \hmax(0)$ as well. Consequently, $\hmax$ has at least one minimizer that resides within $[0,n \mu_{\max}]$.

\subsection{Proof of Lemma~\ref{lem:derivative_H}}  \label{subsec:proof_lem_derivative_H}

Recalling that $H(r,S) = r + \sum_{i \in S} G_i(r)$, it suffices to show that $G_i(r) = \expar{ [X_i - r]^+ }$ is differentiable, with $\frac{ d }{ dr } G_i(r) = -\pr{ X_i \geq r }$. To this end, we argue  that the right-derivative of the latter function exists, showing in particular that $\partial_+ G_i(r) = -\pr{ X_i \geq r }$. A completely symmetrical argument shows that $\partial_- G_i(r) = -\pr{ X_i \geq r }$ as well.

We begin by observing that, since $[X_i - r]^+$ is a non-negative random variable,
\begin{eqnarray*}
\ex{ [X_i - r]^+ } & = & \int_0^{\infty} \pr{ [X_i - r]^+ > x } dx \\
& = & \int_r^{\infty} \pr{ X_i > x } dx \\
& = & \int_r^{\infty} \pr{ X_i \geq x } dx \ , 
\end{eqnarray*}
where the last equality holds since $X_i$ is continuous. Therefore, for every $\Delta \geq 0$,
\begin{eqnarray*}
G_i( r + \Delta ) - G_i( r ) & = & \int_{r+\Delta}^{\infty} \pr{ X_i \geq x } dx - \int_r^{\infty} \pr{ X_i \geq x } dx \\
& = & - \int_r^{r+\Delta} \pr{ X_i \geq x } dx \ .
\end{eqnarray*}
Now, for every $x \in [r,r+\Delta]$, we clearly have $\prpar{ X_i \geq x} \in [\prpar{ X_i \geq r + \Delta}, \prpar{ X_i \geq r}]$, meaning that
\[ -\pr{ X_i \geq r} ~~\leq~~ \frac{ G_i( r + \Delta ) - G_i( r ) }{ \Delta } ~~\leq~~ -\pr{ X_i \geq r + \Delta} \ . \]
Given that $X_i$ is continuous, we know that $\lim_{\Delta \to 0^+} \pr{ X_i \geq r + \Delta} = \pr{ X_i \geq r }$, implying in turn that $\partial_+ G_i(r) = \lim_{\Delta \to 0^+} \frac{ G_i( r + \Delta ) - G_i( r ) }{ \Delta } = \pr{ X_i \geq r }$, as desired.
\section{Adaptivity Gap of \texorpdfstring{$\bs{2}$}{} for General Random Variables} \label{sec:adaptivity_2}

In this section, we consider the adaptive ProbeMax problem in its broadest form, where $X_1, \ldots, X_n$ are general random variables. Our main result consists of establishing an adaptivity gap of at most $2$ in this context, which will be accompanied by an efficient algorithmic approach for explicitly constructing feasible sets whose expected maximum reward matches this adaptivity gap within factor $1 + \eps$.

\subsection{High-level technical overview} \label{subsec:overview_gap_2}

By Lemma~\ref{lem:prop_calH_min},
we know in particular that an optimal solution to problem~\eqref{eqn:definition_U} indeed exists; we make use of $r^* \geq 0$ to designate one such solution. For ease of exposition, the discussion below will be directed toward deriving our adaptivity gap, meaning that it is analytical in nature. The algorithmic implications of this result will be separately presented in Section~\ref{subsec:algorithm_general}.

\paragraph{$\bs{\infty}$-identifiable subsets.}  We say that a subset $S^+ \in {\cal K}$ is $\infty$-right-identifiable with $\hmax$ at $r^*$ when, for any $\eps > 0$, there exists some $\hat{r} \in (r^*,r^*+\eps)$ such that $H(\hat{r},S^+) = \hmax(\hat{r})$.  Clearly, since the collection of subsets ${\cal K}$ is finite, it necessarily contains at least one subset which is $\infty$-right-identifiable at $r^*$. We proceed by presenting an important characterization of such subsets.  Specifically, the next result, whose proof is provided in Section~\ref{subsec:proof_lem_right_H_min}, shows that for any $\infty$-right-identifiable subset $S^+$, the function $H(\cdot,S^+)$ takes its minimum value over the interval $[r^*,\infty)$ at the point $r^*$.

\begin{lemma} \label{lem:right_H_min}
Let $S^+ \in {\cal K}$ be an $\infty$-right-identifiable subset at $r^*$. Then,
\[ \hmax(r^*) ~~=~~ H(r^*,S^+) ~~=~~ \min_{\hat{r} \in [r^*,\infty)} H(\hat{r},S^+) \ . \]
\end{lemma}

Similarly, a subset $S^- \in {\cal K}$ will be called $\infty$-left-identifiable at $r^*$ when, for any $\eps > 0$, there exists some $\hat{r} \in (r^* - \eps,r^*)$ such that $H(\hat{r},S^+) = \hmax(\hat{r})$. Once again, ${\cal K}$ necessarily contains at least one such subset. The following claim provides an analogous characterization of $\infty$-left-identifiable subsets; we omit the proof, as it is nearly-identical to that of Lemma~\ref{lem:right_H_min}.

\begin{lemma} \label{lem:left_H_min}
Let $S^- \in {\cal K}$ be an $\infty$-left-identifiable subset at $r^*$. Then,
\[ \hmax(r^*) ~~=~~ H(r^*,S^-) ~~=~~ \min_{\hat{r} \in (-\infty,r^*]} H(\hat{r},S^-) \ . \]
\end{lemma}

\paragraph{Roots of $\bs{G(r,S) =r}$ and their relation to $\bs{\mybound}$.} Recalling from Section~\ref{subsec:properties_HrS} that $G(r,S) = \sum_{i \in S} G_i(r) = \sum_{i \in S} \expar{ [X_i - r]^+ }$, we argue that the equation $G(r,S) =r$ has a single root, $\rho(S)$, which is non-negative. To verify this claim, one should simply observe that $G(\cdot,S)$ is continuous and weakly-decreasing, with $G(0,S) = \sum_{i \in S} \mu_i$ and $\lim_{r \to \infty} G(r,S) = 0$. The latter explanation concurrently leads to the next claim, which will be useful later on.

\begin{observation} \label{obs:relation_G_r}
$G(r,S) > r$ for all $r < \rho(S)$, and conversely, $G(r,S) < r$ for all $r > \rho(S)$.
\end{observation}

Now, let $S^+$ and $S^-$ be a pair of $\infty$-right-identifiable and $\infty$-left-identifiable subsets at $r^*$, respectively; as previously explained, such subsets indeed exist. Our main structural insight relates $\rho(S^+)$ and $\rho(S^-)$ to the min-max upper bound $\mybound$, as formally stated below.

\begin{lemma} \label{lem:uk_vs_splus_sminus}
$\mybound \leq 2 \cdot \max \{ \rho(S^+), \rho(S^-) \}$.
\end{lemma}
\begin{proof}
Our proof considers two cases, depending on the relation between $r^*$ and $\rho(S^-)$:
\begin{itemize}
    \item {\em When $r^* > \rho(S^-)$}: Here, we claim that $\mybound \leq 2 \rho(S^-)$. To this end, note that
    \begin{eqnarray*}
    \mybound & = & \hmax(r^*) \\
    & = & \min_{\hat{r} \in (-\infty,r^*]} H(\hat{r},S^-) \\
    & \leq & H(\rho(S^-), S^-) \\
    & = & \rho(S^-) + G(\rho(S^-), S^-) \\
    & = & 2 \rho(S^-) \ ,
    \end{eqnarray*}
    where the second equality follows from Lemma~\ref{lem:left_H_min}, and the next inequality holds since $\rho(S^-) \leq r^*$ by the case hypothesis.

    \item {\em When $r^* \leq \rho(S^-)$}: In this case, we observe that
    \begin{eqnarray*}
    G(r^*, S^+) & = & G(r^*, S^-) \\
    & \geq & G(\rho(S^-), S^-) \\
    & = & \rho(S^-) \\
    & \geq & r^* \ ,
    \end{eqnarray*}
    where the first equality holds since $H(r^*, S^+) = \hmax(r^*) = H(r^*, S^-)$, by Lemmas~\ref{lem:right_H_min} and~\ref{lem:left_H_min}, and the first inequality follows by noting that $G(\cdot,S^-)$ is weakly-decreasing and that $r^* \leq \rho(S^-)$. Consequently, we know that $r^* \leq \rho(S^+)$, by Observation~\ref{obs:relation_G_r}. As a result, arguments similar to those of the first case show that
    \begin{eqnarray*}
    \mybound & = & \hmax(r^*) \\
    & = & \min_{\hat{r} \in [r^*,\infty)} H(\hat{r},S^+) \\
    & \leq & H(\rho(S^+), S^+) \\
    & = & 2 \rho(S^+) \ .
    \end{eqnarray*}
\end{itemize}
\end{proof}

\paragraph{Employing an adversarial-order prophet inequality.} Now, let $\tilde{S} \in {\cal K}$ be a subset with $\mybound \leq  2\rho(\tilde{S})$, which is known to exist due to Lemma~\ref{lem:uk_vs_splus_sminus}. In what follows, we conclude our analysis by explaining why the expected maximum reward of this subset satisfies $\expar{ M(\tilde{S}) } \geq \frac{ 1 }{ 2 } \cdot \mybound$. Combined with Theorem~\ref{thm:rel_dk_uk}, which states that $\myadapt^* \leq \mybound$, we immediately infer that $\expar{ M(\tilde{S}) } \geq \frac{ 1 }{ 2 } \cdot \myadapt^*$, meaning that the adaptivity gap in this setting is upper-bounded by $2$.

\begin{lemma}
$\expar{ M(\tilde{S}) } \geq \frac{ 1 }{ 2 } \cdot \mybound$.
\end{lemma}
\begin{proof}
The most relevant result for our particular purposes is the prophet inequality due to \cite{SC84}, attained by her  root-threshold-based stopping policy, which is somewhat less-known than the median-based threshold. Specifically, for any subset $S \subseteq [n]$, recalling that $\rho(S)$ is the unique root of $G(r,S) = r$, consider the threshold policy that inspects the random variables $\{ X_i \}_{i \in S}$ in arbitrary order, and stops when (and if) a value of at least $\rho(S)$ is observed. Letting $T_{ \rho(S) }$ be the random stopping time of this policy, \citeauthor{SC84} proved that the latter has an expected reward of $\expar{ X_{ T_{ \rho(S) } } } \geq \rho(S)$.
Consequently, by employing this threshold policy with the above-mentioned subset $\tilde{S}$ as its input, it indeed follows that
\begin{eqnarray*}
\ex{ M(\tilde{S}) } & \geq & \ex{ X_{ T_{ \rho(\tilde{S}) } } } \\
& \geq & \rho(\tilde{S}) \\
& \geq & \frac{ 1 }{ 2 } \cdot \mybound \ .
\end{eqnarray*}
\end{proof}

\subsection{Algorithmic implications} \label{subsec:algorithm_general}

A close inspection of Section~\ref{subsec:overview_gap_2} reveals that our analysis does not directly lead to an efficient construction, due to two main obstacles. First, even though Lemma~\ref{lem:uk_vs_splus_sminus} informs us that the role of $\tilde{S}$ can be played either by the $\infty$-right-identifiable subset $S^+$ or by the $\infty$-left-identifiable subset $S^-$, it is unclear how to compute such subsets in polynomial time. Second, we have been making repeated use of  the optimal solution $r^*$ to problem~\eqref{eqn:definition_U}, which is generally unknown, and may not even be a rational number. The current section is intended to augment our analysis with a number of additional ideas, with the objective of efficient computation in mind.

\paragraph{Step 1: Narrowing down a minimizer.} Following the discussion in Section~\ref{subsec:properties_HrS}, we assume to have already determined a subinterval $[r^-, r^+] \subseteq [0, n \mu_{\max}]$ that contains at least one minimizer $r^* \in R_{\hmax}$, where $r^+ - r^- = \xi = \frac{ \eps \mu_{\max} }{ 20k }$. As previously explained, one can identify such an interval, for example, via $O( \log ( \frac{ n k }{ \eps } ) )$ iterations of Golden-section search, each requiring a single query of the function $\hmax(\cdot)$.

\paragraph{Step 2: Computing $\bs{\tilde{S}^+}$ and $\bs{\tilde{S}^-}$.}
Let us assume without loss of generality that $G_1( r^+) \geq \cdots \geq G_n( r^+)$. In addition, we make use of $k^- \leq k$ to denote the minimal index for which $G_{k^-}( r^+) = G_k( r^+)$. Similarly, $k^+ \geq k$ will designate the maximal index for which $G_{k^+}( r^+) = G_k( r^+)$. Given these definitions, one can easily verify that, for any subset $S \in {\cal K}$, we have $H( r^+, S) = \hmax( r^+ )$ if and only if $S = \{ 1, \ldots, k^- - 1 \} \cup T$ for some subset $T \subseteq \{ k^-, \ldots, k^+ \}$ of cardinality $k - k^- + 1$. We denote the collection of such subsets by ${\cal F}^+$. Out of all subsets in ${\cal F}^+$, we proceed by computing $\tilde{S}^+$, which is one that maximizes $\sum_{i \in S} G_i( r^+ + \xi )$. It is not difficult to see that, in order to construct $\tilde{S}^+$, on top of choosing $\{ 1, \ldots, k^- - 1 \}$, the $k - k^- + 1$ remaining elements should simply be picked out of $\{ k^-, \ldots, k^+ \}$ in weakly-decreasing order of $G_i( r^+ + \xi)$. By duplicating this construction with respect to $r^-$, we similarly compute a subset $\tilde{S}^- \in {\cal F}^-$ such that $H( r^-, \tilde{S}^-) = \hmax( r^- )$ and such that $\tilde{S}^-$ maximizes $\sum_{i \in S} G_i( r^+ - \xi )$.

\subsection{Analysis} \label{subsec:analysis_approx_splus_sminus}

\paragraph{Properties of $\bs{\tilde{S}^+}$ and $\bs{\tilde{S}^-}$.} The next claim regarding $\tilde{S}^+$ and $\tilde{S}^-$ can be viewed as an approximate analog of Lemmas~\ref{lem:right_H_min} and~\ref{lem:left_H_min}, which will be shown to be sufficient for our purposes. For ease of presentation, the proof of this result is deferred to Section~\ref{subsec:proof_lem_approx_splus_sminus}.

\begin{lemma} \label{lem:approx_splus_sminus}
The subsets $\tilde{S}^+$ and $\tilde{S}^-$ satisfy the following properties:
\begin{enumerate}
    \item $\hmax( r^+ ) \leq \min_{\hat{r} \in [r^+,\infty)} H( \hat{r},\tilde{S}^+) + k \xi$.

    \item $\hmax( r^- ) \leq \min_{\hat{r} \in (-\infty,r^-]} H( \hat{r},\tilde{S}^-) + k \xi$.
\end{enumerate}
\end{lemma}

\paragraph{Deriving the adaptivity gap.} Out of the subsets $\tilde{S}^+$ and $\tilde{S}^-$ we have just constructed, let $\tilde{S}$ be the one whose $\rho(\cdot)$ value is maximized. Mimicking the closing discussion of Section~\ref{subsec:overview_gap_2}, in order to prove that $\expar{ M(\tilde{S}) } \geq ( \frac{ 1 }{ 2 } - \eps ) \cdot \myadapt^*$, it suffices to show that $\mybound \leq (2 + \eps) \cdot \rho( \tilde{S} )$, which is precisely what Lemma~\ref{lem:rel_uk_tildes_pm} below accomplishes. Indeed, by employing the stopping policy of \cite{SC84} with $\rho( \tilde{S} )$ as our threshold, we observe that the maximum reward $M(\tilde{S})$ of this set has an expected value of
\[ \ex{ M(\tilde{S}) } ~~\geq~~ \rho(\tilde{S}) ~~\geq~~ \frac{ 1 }{ 2 + \eps } \cdot \mybound ~~\geq~~ \left( \frac{ 1 }{ 2 } - \eps \right) \cdot \myadapt^* \ . \]

\begin{lemma} \label{lem:rel_uk_tildes_pm}
$\mybound \leq (2 + \eps) \cdot \max \{ \rho( \tilde{S}^+) , \rho( \tilde{S}^-) \}$.
\end{lemma}
\begin{proof}
We consider two scenarios, based on the relation between $r^-$ and $\rho(\tilde{S}^-)$:
\begin{itemize}
    \item {\em When ${r^- > \rho(\tilde{S}^-)}$}: In this case, we claim that $\mybound \leq (2 + \eps) \cdot  \rho(\tilde{S}^-)$. For this purpose, since $r^*$ is an optimal solution to problem~\eqref{eqn:definition_U}, we have
    \begin{eqnarray*}
    \mybound & = & \hmax(r^*)  \\
    & \leq & \hmax( r^- ) \\
    & \leq& \min_{\hat{r} \in (-\infty,r^-]} H( \hat{r},\tilde{S}^-) + k \xi \\
    & \leq & H(\rho(\tilde{S}^-), \tilde{S}^-) + \frac{ \eps \mu_{\max} }{ 20 }  \\
    & = & 2 \rho(\tilde{S}^-) + \frac{ \eps \mu_{\max} }{ 20 } \ .
    \end{eqnarray*}
    Here, the second inequality follows from item~2 of Lemma~\ref{lem:approx_splus_sminus}, and the third inequality is obtained by recalling that $\rho(\tilde{S}^-) < r^-$ and that $\xi = \frac{ \eps \mu_{\max} }{ 20 k }$. By rearranging the inequality above, it follows that
    \begin{eqnarray*}
    \mybound & \leq & 2 \cdot \left( 1 - \frac{ \eps \mu_{\max} }{ 20 \mybound } \right)^{-1} \cdot \rho(\tilde{S}^-) \\
    & \leq & 2 \cdot \left( 1 - \frac{ \eps }{ 20 } \right)^{-1} \cdot  \rho(\tilde{S}^-) \\
    & \leq & (2 + \eps) \cdot  \rho(\tilde{S}^-) \ .
    \end{eqnarray*}
    To better understand the second inequality, note that the optimum adaptive reward $\myadapt^*$ is clearly lower-bounded by the maximal expectation, $\mu_{\max}$; in conjunction with Theorem~\ref{thm:rel_dk_uk}, it follows that $\mybound \geq \myadapt^* \geq \mu_{\max}$.

    \item {\em When ${r^- \leq \rho(\tilde{S}^-)}$}: In this case, we show that $\mybound \leq (2 + \eps) \cdot  \rho(\tilde{S}^+)$. To this end, we first present an auxiliary claim, showing that $r^+$ cannot be much larger than $\rho(\tilde{S}^+)$ whenever $r^- \leq \rho(\tilde{S}^-)$. We defer the proof to Section~\ref{subsec:proof_clm_aux_case2_approx}.

    \begin{claim} \label{clm:aux_case2_approx}
    $r^+ \leq \rho(\tilde{S}^+) + (k+2) \cdot \xi$.
    \end{claim}

    Given this result, we observe that
    \begin{eqnarray}
    \mybound & = & \hmax(r^*) \nonumber \\
    & \leq & \hmax( r^+ ) \nonumber \\
    & \leq & \min_{\hat{r} \in [r^+,\infty)} H( \hat{r},\tilde{S}^+) + k \xi \label{eqn:proof_alg_2bound_case2_2} \\
    & \leq & H(\rho(\tilde{S}^+) + (k+2) \cdot \xi, \tilde{S}^+) + k \xi \label{eqn:proof_alg_2bound_case2_3} \\
    & \leq & H(\rho(\tilde{S}^+), \tilde{S}^+) + 2(k+1) \cdot \xi \nonumber \\
    & \leq & 2 \rho(\tilde{S}^+) + \frac{ \eps \mu_{\max} }{ 5 } \ , \nonumber
    \end{eqnarray}
    where inequality~\eqref{eqn:proof_alg_2bound_case2_2} follows from item~1 of Lemma~\ref{lem:approx_splus_sminus}, and inequality~\eqref{eqn:proof_alg_2bound_case2_3} is implied by Claim~\ref{clm:aux_case2_approx}. Now, by rearranging the inequality above, it follows that
    \begin{eqnarray*}
    \mybound & \leq & 2 \cdot \left( 1 - \frac{ \eps \mu_{\max} }{ 5\mybound } \right)^{-1} \cdot \rho(\tilde{S}^+) \\
    & \leq & 2 \cdot \left( 1 - \frac{\eps}{5} \right)^{-1} \cdot  \rho(\tilde{S}^+) \\
    & \leq & (2 + \eps) \cdot  \rho(\tilde{S}^+) \ .
    \end{eqnarray*}
\end{itemize}
\end{proof}

\subsection{Proof of Lemma~\ref{lem:right_H_min}} \label{subsec:proof_lem_right_H_min}
First, since $S^+$ is $\infty$-right-identifiable at $r^*$, it follows that there is a sequence of points $r_1 > r_2 > \cdots$ such that $\lim_{t \to \infty} r_t = r^*$ and such that $H( r_t, S^+ ) = \hmax( r_t )$ for all $t \geq 1$. By Observation~\ref{obs:prop_H2_min} and Lemma~\ref{lem:prop_calH_min}, we know that both $H(\cdot, S^+)$ and $\hmax(\cdot)$ are convex, and therefore continuous, implying that
\[ H(r^*,S^+) ~~=~~ \lim_{t \to \infty} H( r_t, S^+ ) ~~=~~ \lim_{t \to \infty} \hmax( r_t ) ~~=~~ \hmax(r^*) \ , \]
which is precisely the first equality we wish to establish.

Now, to prove that $H(r^*,S^+) = \min_{\hat{r} \in [r^*,\infty)} H(\hat{r},S^+)$, suppose by way of contradiction that there exists some $\bar{r} > r^*$ for which $H(\bar{r},S^+) < H(r^*,S^+)$. As a result, since $H(\cdot, S^+)$ is convex by Observation~\ref{obs:prop_H2_min}, it follows that for every $\lambda \in (0,1)$ we have
\begin{eqnarray*}
H( \lambda \cdot r^* + (1 - \lambda) \cdot \bar{r}, S^+ ) & \leq & \lambda \cdot H(r^*, S^+) + (1 - \lambda) \cdot H(\bar{r}, S^+ ) \\
& < & H(r^*, S^+ ) \\
& = & \hmax(r^*) \\
& \leq & \hmax( \lambda \cdot r^* + (1 - \lambda) \cdot \bar{r} ) \ ,
\end{eqnarray*}
where the last inequality holds since $r^*$ is a minimizer of $\hmax(\cdot)$. In other words, we have just shown that $H(\hat{r}, S^+) < \hmax( \hat{r} )$ for all $\hat{r} \in (r^*, \bar{r})$, contradicting the fact that $S^+$ is $\infty$-right-identifiable with $\hmax$ at $r^*$.

\subsection{Proof of Lemma~\ref{lem:approx_splus_sminus}} \label{subsec:proof_lem_approx_splus_sminus}

In what follows, we prove item~1, stating that $\hmax( r^+ ) \leq \min_{\hat{r} \in [r^+,\infty)} H( \hat{r},\tilde{S}^+) + k \xi$. The proof of item~2, which is the analogous claim with respect to $\tilde{S}^-$, proceeds along nearly identical arguments, and is therefore omitted. To prove the desired upper bound on $\hmax( r^+ )$, let $\bar{r}$ be a minimizer of $H(\cdot, \tilde{S}^+)$ over the interval $[r^+,\infty)$. Our proof considers two cases, depending on the relation between $r^+$ and $\bar{r}$:
\begin{itemize}
    \item {\em When ${\bar{r} \in [r^+, r^+ + \xi]}$}: In this case, by construction of $\tilde{S}^+$, we have
    \begin{eqnarray*}
    \hmax(r^+) & = & H(r^+, \tilde{S}^+)  \\
    & = & r^+ + \sum_{i \in \tilde{S}^+} \expar{ [X_i - r^+]^+ } \\
    & \leq & \bar{r} + \sum_{i \in \tilde{S}^+} \expar{ [X_i - \bar{r}]^+ } + k \xi \\
    & = & H(\bar{r}, \tilde{S}^+) + k \xi \ .
    \end{eqnarray*}
    where the inequality above follows from the case hypothesis, ${\bar{r} \in [r^+, r^+ + \xi]}$.

    \item {\em When ${\bar{r} \in (r^+ + \xi, \infty)}$}: Here, the crucial observation is that our proof of Lemma~\ref{lem:right_H_min} does not assume that $r^*$ is a minimizer of $\hmax( \cdot )$ at any time, meaning that it applies to any point in $\bbR$. In particular, precisely the same arguments show that, letting $S^+ \in {\cal K}$ be an $\infty$-right-identifiable subset at $r^+$, we have
    \begin{equation} \label{eqn:comp_rplus_rstar_0}
    \hmax(r^+) ~~=~~ H(r^+,S^+) ~~=~~ \min_{\hat{r} \in [r^+,\infty)} H(\hat{r},S^+) \ . \end{equation}
    On the other hand,
    \begin{eqnarray}
    H(r^+ + \xi, \tilde{S}^+) & = & r^+ + \xi + \sum_{i \in \tilde{S}^+} G_i( r^+ + \xi ) \nonumber \\
    & \geq & r^+ + \xi + \sum_{i \in {S}^+} G_i( r^+ + \xi ) \label{eqn:proof_lem_approx_splus_sminus_eq4} \\
    & = & H(r^+ + \xi, {S}^+) \nonumber \\
    & \geq & \min_{\hat{r} \in [r^+,\infty)} H(\hat{r},S^+) \nonumber  \\
    & = & \hmax(r^+) \label{eqn:proof_lem_approx_splus_sminus_eq5} \\
    & = & H( r^+, \tilde{S}^+ ) \ . \label{eqn:proof_lem_approx_splus_sminus_eq6}
    \end{eqnarray}
    In this case, inequality~\eqref{eqn:proof_lem_approx_splus_sminus_eq4} follows by recalling that $\tilde{S}^+$ is a subset that maximizes $\sum_{i \in S} G_i( r^+ + \xi )$ over ${\cal F}^+$, by construction. However, equation~\eqref{eqn:comp_rplus_rstar_0} implies that $S^+ \in {\cal F}^+$ as well, and therefore $\sum_{i \in \tilde{S}^+} G_i( r^+ + \xi ) \geq \sum_{i \in {S}^+} G_i( r^+ + \xi )$. Equality~\eqref{eqn:proof_lem_approx_splus_sminus_eq5} is precisely~\eqref{eqn:comp_rplus_rstar_0}. Finally, equality~\eqref{eqn:proof_lem_approx_splus_sminus_eq6} holds since, as mentioned earlier,   $\tilde{S}^+ \in {\cal F}^+$ by construction.

    Therefore, since $H(\cdot, \tilde{S}^+)$ is convex by Observation~\ref{obs:prop_H2_min}, this function must be weakly-increasing over $[r^+ + \xi, \infty)$, implying in particular that $H(\bar{r} , \tilde{S}^+) \geq H(r^+ , \tilde{S}^+)$, since ${\bar{r} \in (r^+ + \xi, \infty)}$ by the case hypothesis. Combined with equality~\eqref{eqn:proof_lem_approx_splus_sminus_eq6}, it  follows that $\hmax( r^+ ) \leq H(\bar{r} , \tilde{S}^+)$.
\end{itemize}

\subsection{Proof of Claim~\ref{clm:aux_case2_approx}} \label{subsec:proof_clm_aux_case2_approx}

Suppose by contradiction that $r^+ > \rho(\tilde{S}^+) + (k+2) \cdot \xi$. In what follows, we show that the latter inequality implies $G( r^-, \tilde{S}^- ) < r^-$. As a result, by Observation~\ref{obs:relation_G_r}, we must have $r^- > \rho( \tilde{S}^- )$, contradicting our case hypothesis, $r^- \leq \rho(\tilde{S}^-)$.

To this end, since $r^+ > \rho(\tilde{S}^+) + (k+2) \cdot \xi$, we first observe that
\begin{eqnarray}
G(r^+, \tilde{S}^+) & \leq & G(\rho(\tilde{S}^+), \tilde{S}^+) \nonumber\\
& = & \rho(\tilde{S}^+) \nonumber \\
& < & r^+ - (k+2) \cdot  \xi \ , \label{eqn:bound_gplus_aux_claim}
\end{eqnarray}
where the first inequality holds since $G(\cdot, \tilde{S}^+)$ is weakly-decreasing. Given this result, we proceed to show that $G( r^-, \tilde{S}^- ) < r^-$ by noting that
\begin{eqnarray}
G( r^-, \tilde{S}^- ) & = & H( r^-, \tilde{S}^- ) - r^- \nonumber \\
& = & \hmax( r^- ) - r^- \label{subsec:proof_clm_aux_case2_approx_eq1}\\
& \leq & \hmax( r^+ ) - r^+ + (k+1) \cdot \xi \label{subsec:proof_clm_aux_case2_approx_eq4} \\
& = & H( r^+, \tilde{S}^+ ) - r^+ + (k+1) \cdot \xi \label{subsec:proof_clm_aux_case2_approx_eq2} \\
& = & G( r^+, \tilde{S}^+ ) + (k+1) \cdot \xi \nonumber \\
& < & r^+ - \xi \label{subsec:proof_clm_aux_case2_approx_eq3} \\
& \leq & r^- \ . \nonumber
\end{eqnarray}
Here, equalities~\eqref{subsec:proof_clm_aux_case2_approx_eq1} and~\eqref{subsec:proof_clm_aux_case2_approx_eq2} hold, respectively, since $\tilde{S}^- \in {\cal F}^-$ and $\tilde{S}^+ \in {\cal F}^+$, whereas inequality~\eqref{subsec:proof_clm_aux_case2_approx_eq3} follows from~\eqref{eqn:bound_gplus_aux_claim}. To better understand the main transition, inequality~\eqref{subsec:proof_clm_aux_case2_approx_eq4}, note that
\begin{eqnarray*}
\hmax( r^+ ) & \geq & H( r^+, \tilde{S}^- ) \\
& = & r^+ + \sum_{i \in \tilde{S}^-} \ex{ [X_i - r^+]^+ } \\
& \geq & r^- + \sum_{i \in \tilde{S}^-} \ex{ [X_i - r^-]^+ } - k \xi \\
& = & H( r^-, \tilde{S}^-) - k \xi \\
& = & \hmax( r^- ) - k \xi \ .
\end{eqnarray*}
\section{Adaptivity Gap of \texorpdfstring{$\bs{\frac{ e }{ e -1 }}$}{} for Continuous Random Variables} \label{sec:adapt_gap_e_em1}

In what follows, we examine a more lenient formulation of the adaptive ProbeMax problem, where $X_1, \ldots, X_n$ are assumed to be continuous random variables. In this setting, we establish an improved adaptivity gap of $\frac{ e }{ e -1 } \approx 1.58$, along with an efficient construction of feasible sets whose expected maximum reward matches this gap. Moving forward, we dedicate Section~\ref{subsec:structure_thm} to introducing a linear extension of problem~\eqref{eqn:definition_U} and to proving that the latter admits highly-structured optimal solutions. Then, Sections~\ref{subsec:prop_policy_e_em1} and~\ref{subsec:cont_analysis_overview} will provide a high-level overview of our probing policy and its analysis, leaving most proofs to be presented in subsequent sections.

\subsection{The linear extension and its optimal solutions} \label{subsec:structure_thm}

We remind the reader that, in Theorem~\ref{thm:rel_dk_uk}, we have shown that
\begin{equation} \label{eqn:definition_U_rep} \tag{\ref{eqn:definition_U} revisited}
\mybound ~~=~~ \min_{r \in \bbR} \max_{S \in {\cal K}} H(r,S) \ .
\end{equation}
provides an upper bound on the adaptive optimum $\myadapt^*$, noting that $H(r,S) = r + \sum_{i \in S} G_i(r)$. Now, let us consider a continuous relaxation of this problem where, instead of maximizing over the collection of subsets ${\cal K}$, we allow fractional solutions within the set $\Psi = \{ \psi \in [0,1]^n : \| \psi \|_1 = k \}$. Formally speaking, we are focusing our attention on 
\begin{equation} \label{eqn:prob_min_max_cont}
\min_{r \in \bbR} \max_{\psi \in \Psi} \bar{H}(r, \psi) \ , \tag{$\overline{\text{MinMax}}$}
\end{equation}
where the function $\bar{H}(\cdot,\cdot)$ is a linear extension of $H(\cdot,\cdot)$ from $\bbR \times {\cal K}$ to $\bbR \times \Psi$, given by
\[ \bar{H}(r,\psi) ~~=~~ r + \sum_{i \in [n]} G_i(r) \cdot \psi_i \ . \]
The next claim, whose proof is given in Section~\ref{subsec:proof_lem_min_max_cont_disc}, argues that the optimal values of both problems are actually identical.

\begin{lemma} \label{lem:min_max_cont_disc}
$\opt\eqref{eqn:prob_min_max_cont} = \opt\eqref{eqn:definition_U}$.
\end{lemma}

\paragraph{The structure theorem.} Let $r^*$ be an optimal solution to problem~\eqref{eqn:prob_min_max_cont}. As shown within the proof of Lemma~\ref{lem:min_max_cont_disc}, given this point, its corresponding problem $\max_{\psi \in \Psi} \bar{H}(r^*, \psi)$ admits an integer optimal solution. However, we do not know whether such solutions are useful in improving on the adaptivity gap of $2$, which is applicable to general random variables, as explained in Section~\ref{sec:adaptivity_2}. That said, we are currently considering the setting where $X_1, \ldots, X_n$ are continuous, in which case each of the functions $\{ H( \cdot, S) \}_{S \in {\cal K}}$ was shown in Lemma~\ref{lem:derivative_H} to be differentiable. In the remainder of this section, we exploit the special form of these derivatives to argue that $\max_{\psi \in \Psi} \bar{H}(r^*, \psi)$ admits an almost-integer optimal solution $\psi^*$ satisfying $\sum_{i \in [n]} \prpar{ X_i \geq r^*} \cdot \psi_i^* = 1$. This structural property, whose upcoming proof is based on a polynomial-time explicit construction, will become crucial later on. 

\begin{theorem} \label{thm:structure_psi_star}
There exists an optimal solution $\psi^*$ to $\max_{\psi \in \Psi} \bar{H}(r^*, \psi)$ that satisfies the following properties:
\begin{enumerate}
    \item $\sum_{i \in [n]} \prpar{ X_i \geq r^*} \cdot \psi_i^* = 1$.
    
    \item $\psi^*$ has at most two fractional coordinates.
\end{enumerate}
\end{theorem}

As a side note, from this point on, we will be directly working with the exact value of the minimizer $r^*$, rather than with a tiny interval $[r^-, r^+]$ containing $r^*$, as in Section~\ref{subsec:algorithm_general}. This simplifying assumption is made in order to provide the cleanest presentation possible, as opposed to dragging lower-order terms throughout our analysis.

\paragraph{Step 1: Initial construction of $\bs{S^-}$ and $\bs{S^+}$.} Let us assume without loss of generality that $G_1( r^* ) \geq \cdots \geq G_n(r^*)$. Adopting some of the notation introduced in Section~\ref{subsec:algorithm_general}, we make use of $k^- \leq k$ to denote the minimal index for which $G_{k^-}(r^*) = G_k(r^*)$. Similarly, $k^+ \geq k$ will designate the maximal index for which $G_{k^+}(r^*) = G_k(r^*)$. Given these definitions, one can easily verify that, for any subset $S \in {\cal K}$, we have $H(r^*, S) = \hmax(r^* )$ if and only if $S = \{ 1, \ldots, k^- - 1 \} \cup T$ for some subset $T \subseteq \{ k^-, \ldots, k^+ \}$ of cardinality $k - k^- + 1$. This collection of subsets will be denoted by ${\cal F}^*$. We proceed to construct a pair of subsets $S^-$ and $S^+$ as follows:
\begin{itemize}
    \item Let $S^-$ be a subset that maximizes $\frac{ d }{ dr } H(r^*,S)$ over ${\cal F}^*$, noting that the latter derivative indeed exists, by Lemma~\ref{lem:derivative_H}. Moreover, since $\frac{ d }{ dr } H(r^*,S) = 1 - \sum_{i \in S} \pr{ X_i \geq r^* }$, we can efficiently compute $S^-$ by first picking $\{ 1, \ldots, k^- - 1 \}$, and then adding $k - k^- + 1$ elements out of $\{ k^-, \ldots, k^+ \}$ by weakly-increasing order of $\prpar{ X_i \geq r^* }$.
    
    \item Similarly, $S^+$ is a subset that minimizes $\frac{ d }{ dr } H(r^*,S)$ over ${\cal F}^*$. The construction of $S^+$ is symmetrical to that of $S^-$, with the exception of picking elements out of $\{ k^-, \ldots, k^+ \}$ by weakly-decreasing order of $\prpar{ X_i \geq r^* }$.
\end{itemize}
The next result, whose proof appears in Section~\ref{subsec:proof_lem_prob_Sm_Sp}, informs us that the functions $H(\cdot,S^-)$ and $H(\cdot,S^+)$ respectively have non-negative and non-positive derivatives at $r^*$.

\begin{lemma} \label{lem:prob_Sm_Sp}
$\frac{ d }{ dr } H(r^*,S^-) \geq 0$ and $\frac{ d }{ dr } H(r^*,S^+) \leq 0$.
\end{lemma}

\paragraph{Step 2: Maximizing the overlap between $\bs{S^-}$ and $\bs{S^+}$.} We move on to argue that the subsets $S^-$ and $S^+$ can be assumed to be overlapping in at least $k-1$ elements. To attain this property, when $| S^- \cap S^+ | \leq k-2$, let us pick two arbitrary elements, $i^- \in S^- \setminus S^+$ and $i^+ \in S^+ \setminus S^-$, noting that by the preceding discussion, both must reside within $\{ k^-, \ldots, k^+ \}$. In addition, let $\hat{S} = (S^+ \setminus \{ i^+ \}) \cup \{ i^- \}$ be the subset obtained by swapping $i^-$ into $S^+$ in place of $i^+$; clearly, $\hat{S} \in {\cal F}^*$ as well. We proceed by considering two cases:
\begin{itemize}
    \item {\em When $\frac{ d }{ dr } H(r^*,\hat{S}) \geq 0$:} Here, $\hat{S}$ and $S^+$ are two sets in ${\cal F}^*$ with $\frac{ d }{ dr } H(r^*,\hat{S}) \geq 0$ and $\frac{ d }{ dr } H(r^*,S^+) \leq 0$. Moreover, $| \hat{S} \cap S^+ | = k-1$, meaning that by renaming $\hat{S}$ as $S^-$, we are done.
    
    \item {\em When $\frac{ d }{ dr } H(r^*,\hat{S}) < 0$:} In this case, $S^-$ and $\hat{S}$ are two sets in ${\cal F}^*$ with $\frac{ d }{ dr } H(r^*,S^-) \geq 0$ and $\frac{ d }{ dr } H(r^*,\hat{S}) < 0$. Moreover, $| S^- \cap \hat{S} | = | S^- \cap S^+ | + 1$, meaning that by renaming $\hat{S}$ as $S^+$, the overlap between $S^-$ and $S^+$ increases by another element. 
\end{itemize}
This swapping procedure can be reiterated until we end up with $| S^- \cap S^+ | \geq k-1$.

\paragraph{Step 3: Defining $\bs{\psi^*}$.} By Lemma~\ref{lem:derivative_H}, we know that $\frac{ d }{ dr } H(r^*,S) = 1 - \sum_{i \in S} \pr{ X_i \geq r^* }$ for every set $S \subseteq [n]$. Therefore, in light of Lemma~\ref{lem:prob_Sm_Sp}, we have $\sum_{i \in S^-} \prpar{ X_i \geq r^* } \leq 1$ and $\sum_{i \in S^+} \prpar{ X_i \geq r^* } \geq 1$, implying that there exists some $\alpha \in [0,1]$ for which $\alpha \cdot \sum_{i \in S^+} \prpar{ X_i \geq r^*} + (1 - \alpha) \cdot \sum_{i \in S^-} \prpar{ X_i \geq r^*} = 1$. Given this value, let us define $\psi^* = \alpha \cdot \chi^{S^+} + (1 - \alpha) \cdot \chi^{S^-}$, where $\chi^{S^+}$ and $\chi^{S^-}$ are the characteristic vectors of $S^+$ and $S^-$, respectively. 

\paragraph{Analysis.} We conclude our analysis by showing that $\psi^*$ indeed satisfies the structural properties required by Theorem~\ref{thm:structure_psi_star}. First, to derive property~1, we observe that 
\begin{equation} \label{eqn:prop_psistar_eq1}
\sum_{i \in [n]} \pr{ X_i \geq r^*} \cdot \psi_i^* ~~=~~ \alpha \cdot \sum_{i \in S^+} \pr{ X_i \geq r^*} + (1 - \alpha) \cdot \sum_{i \in S^-} \pr{ X_i \geq r^*} ~~=~~ 1 \ ,
\end{equation}
where the second equality follows from the above-mentioned choice of $\alpha$. In addition, as explained in step~2, the sets $S^-$ and $S^+$ overlap in at least $k-1$ elements, implying that the vector 
$\psi^* = \alpha \cdot \chi^{S^+} + (1 - \alpha) \cdot \chi^{S^-}$ has at most two fractional coordinates. Finally, the next claim, whose proof appears in Section~\ref{subsec:proof_lem_F_rstar_psistar}, shows that $\psi^*$ constitutes an optimal solution to $\max_{\psi \in \Psi} \bar{H}(r^*, \psi)$.

\begin{lemma} \label{lem:F_rstar_psistar}
$\bar{H}(r^*, \psi^*) = \max_{\psi \in \Psi} \bar{H}(r^*, \psi)$.
\end{lemma}

\subsection{The probing policy} \label{subsec:prop_policy_e_em1}

With the vector $\psi^*$ in hand, by recalling that $\| \psi^* \|_1 = k$, we observe that $\psi^*$ has either two fractional coordinates or none, by Theorem~\ref{thm:structure_psi_star}. From this point on, we consider the former case, letting $\ell$ and $m$ be these two coordinates. The case where $\psi^*$ is integer-valued can be addressed through precisely the same analysis, with minor simplifications along the way.

\paragraph{Notation.} For convenience of notation, we introduce the random variable $W \sim \mybern( \psi^*_{ \ell })$, which is independent of $X_1, \ldots, X_n$, and let $X_{\ell,m} = W X_{\ell} + (1-W) X_m$. In addition, we rename the collection of random variables $\{ X_i : \psi^*_i = 1, i \in [n] \} \cup \{ X_{\ell,m} \}$ as $Y_1, \ldots, Y_k$. To avoid confusion between different indices, the mapping $\pi : [k] \to [n]$ will match between $Y$ and $X$ variables, in the sense that $Y_i$ is our renaming for $X_{\pi(i)}$. Furthermore, we use $\hat{i} \in [k]$ to denote the $Y$-index corresponding to $X_{\ell,m}$, meaning that $Y_{\hat{i}} = X_{\ell,m}$. 

\paragraph{The free-order probing policy.} To describe our policy, let us assume without loss of generality that
\[ \ex{ Y_1 | Y_1 \geq r^* } ~~\geq~~ \cdots ~~\geq~~ \ex{ Y_k | Y_k \geq r^* } \ . \]
Then, we inspect the sequence of random variables $Y_1, \ldots, Y_k$ in this particular order, and stop at the first sample whose value is at least $r^*$, when such a sample exists. As an aside, we refer the reader's attention to the fact that, due to the randomness in $W$, we are not operating on a deterministic set of $k$ random variables out of $\{ X_i \}_{i \in [n]}$, but rather on $\{ X_i : \psi^*_i = 1, i \in [n] \}$ and $X_{\ell,m}$, which is a mixture of $X_{\ell}$ and $X_m$. In the sequel, we explain how our policy can easily be derandomized. 

\subsection{High-level analysis} \label{subsec:cont_analysis_overview}

\paragraph{The expected reward.} In order to express the expected reward of our policy, we first introduce the indicator random variable $B_i = \mathbbm{1}[ Y_i \geq r^* ]$ for every $i \in [k]$. Clearly, $B_1, \ldots, B_k$ are mutually independent, and we denote their combined value by $B = \sum_{i \in [k]} B_i$. The next two claims, whose respective proofs appear in Sections~\ref{subsec:proof_lem_e_e-1_exp_B} and~\ref{subsec:proof_lem_e_e-1_exp_min_binom} will be useful later on.

\begin{lemma} \label{lem:e_e-1_exp_B}
$\expar{ B } = 1$.
\end{lemma}

\begin{lemma} \label{lem:e_e-1_exp_min_binom}
$\expar{ \min \{ B, 1 \} } \geq 1 - \frac{ 1 }{ e }$.
\end{lemma}

Now, let $T$ be the random stopping time of our policy. That is, when $B_i = 1$ for at least one index $i \in [k]$, the stopping time $T$ corresponds to the minimal such index. Otherwise, our probing policy does not stop, which will be indicated by $T = \infty$. In other words,
\[ T ~~=~~ \begin{cases}
    \min\{ i \in [k] : B_i = 1 \}, \qquad & \text{when } B \geq 1 \\
     \infty, & \text{when } B = 0
     \end{cases} \]
With this notation, we obtain the random reward $Y_T$, with the convention that $Y_{\infty} = 0$. Therefore,
\begin{equation} \label{eq:exp_R1}
\ex{ Y_T } ~~=~~ \sum_{t \in [k]} \pr{ T = t } \cdot \ex{ Y_T | T = t } ~~=~~ \sum_{t \in [k]} \pr{ T = t } \cdot \ex{ Y_t | Y_t \geq r^* } \ ,
\end{equation}
where the second equality holds since $Y_1, \ldots, Y_k$ are independent.

\paragraph{Upper-bounding method.} Toward relating this quantity to the adaptive optimum $\myadapt^*$, we introduce an intermediate random variable $Y^{ \Sigma }$, standing for the total reward attained by inspecting the sequence $Y_1, \ldots, Y_k$, and picking each and every sample whose value is at least $r^*$. In other words, $Y^{ \Sigma } = \sum_{i \in [k]} B_i Y_i$. As such,
\begin{equation} \label{eq:exp_Rinfty_1}
\ex{ Y^{ \Sigma } } ~~=~~ \sum_{i \in [k]} \ex{ B_i Y_i } ~~=~~ \sum_{i \in [k]} \pr{ B_i = 1 } \cdot \ex{ Y_i | Y_i \geq r^* } \ .
\end{equation}
Yet another representation of $Y^{ \Sigma }$ is via the stopping time $T$. For this purpose, let $Y^{ \Sigma }_{ [t,k] }$ be the random total reward accumulated over $Y_t, \ldots, Y_k$, namely, $Y^{ \Sigma }_{ [t,k] } = \sum_{i \in [t,k]} B_i Y_i$. Then, 
\begin{equation} \label{eq:exp_Rinfty_2}
Y^{ \Sigma } ~~=~~ Y_T + Y^{ \Sigma }_{ [T+1,k] } \ .
\end{equation}
In Sections~\ref{subsec:proof_lem_e_e-1_exp_Rinfty}  and~\ref{subsec:proof_lem_e_e-1_exp_diff}, we prove the next two claims, connecting between $\expar{ Y^{ \Sigma } }$, $\expar{ Y_T  }$, and our min-max bound $\mybound$.

\begin{lemma} \label{lem:e_e-1_exp_Rinfty}
$\expar{ Y^{ \Sigma } } = \mybound$.
\end{lemma}

\begin{lemma} \label{lem:e_e-1_exp_diff}
$\expar{ Y^{ \Sigma } } - \expar{ Y_T  } \leq \expar{ [B-1]^+ } \cdot \mybound$.
\end{lemma}

\paragraph{Relating $\bs{\expar{ Y_T }}$ to the adaptive optimum $\bs{\myadapt^*}$.} We conclude by lower-bounding the expected reward $\expar{ Y_T }$ of our policy in terms of the adaptive optimum $\myadapt^*$, going through the min-max bound $\mybound$ as follows:
\begin{eqnarray}
\ex{ Y_T } & \geq & \ex{ Y^{ \Sigma } } - \ex{ [B-1]^+ } \cdot \mybound \label{eqn:rel_YT_adapt_opt_1} \\
& = & \left( 1 - \ex{ [B-1]^+ } \right) \cdot \mybound \label{eqn:rel_YT_adapt_opt_2} \\
& = & \left( \ex{B} - \ex{ [B-1]^+ } \right) \cdot \mybound \label{eqn:rel_YT_adapt_opt_3} \\
& = & \ex{ \min \{ B, 1 \} } \cdot \mybound \nonumber \\
& \geq & \left( 1 - \frac{ 1 }{ e } \right) \cdot \mybound \label{eqn:rel_YT_adapt_opt_4} \\
& \geq & \left( 1 - \frac{ 1 }{ e } \right) \cdot \myadapt^* \label{eqn:rel_YT_adapt_opt_5} \ . 
\end{eqnarray}
Here, the first three transitions, \eqref{eqn:rel_YT_adapt_opt_1}-\eqref{eqn:rel_YT_adapt_opt_3}, follow from Lemmas~\ref{lem:e_e-1_exp_diff}, \ref{lem:e_e-1_exp_Rinfty}, and~\ref{lem:e_e-1_exp_B}, respectively, whereas inequality~\eqref{eqn:rel_YT_adapt_opt_4} follows from Lemma~\ref{lem:e_e-1_exp_min_binom}. Finally, inequality~\eqref{eqn:rel_YT_adapt_opt_5} holds since $\mybound$ forms an upper bound on $\myadapt^*$, as shown in Theorem~\ref{thm:rel_dk_uk}.

\paragraph{Derandomization.} A close inspection of the sequence of inequalities we have just established reveals that it falls slightly short of constructing a single set of $k$ random variables to be non-adaptively probed. Specifically, as mentioned in Section~\ref{subsec:prop_policy_e_em1}, our policy does not operate on a deterministic set of $k$ random variables out of $\{ X_i \}_{i \in [n]}$, but rather on $\{ X_i : \psi^*_i = 1, i \in [n] \}$ and on the mixture $X_{\ell,m} = W X_{\ell} + (1-W) X_m$, where $W \sim \mybern( \psi^*_{ \ell })$ is sampled independently of $X_1, \ldots, X_n$.

That said, given the very simple way in which $X_{\ell,m}$ is defined, we can easily derandomize this policy. Specifically, when $\expar{ Y_T | W=1 } \geq \expar{Y_T}$, we employ our policy on $\{ X_i : \psi^*_i = 1, i \in [n] \} \cup \{ X_{\ell} \}$, while preserving its inspection order; in the opposite case, our policy is employed on $\{ X_i : \psi^*_i = 1, i \in [n] \} \cup \{ X_m \}$. Consequently, letting $\tilde{S}$ be the resulting set of the $k$ random variables to be probed, we infer that the maximum reward $M(\tilde{S})$ has an expected value of  
\begin{eqnarray*}
\ex{ M(\tilde{S}) } & \geq & \max \left\{ \ex{ Y_T | W=1 }, \ex{ Y_T | W=0 } \right\} \\
& \geq & \pr{W=1} \cdot \ex{ Y_T | W=1 } + \pr{W=0} \cdot \ex{ Y_T | W=0 } \\
& = & \ex{Y_T} \\
& \geq & \left( 1 - \frac{ 1 }{ e } \right) \cdot \myadapt^* \ . 
\end{eqnarray*}
It is important to note that each of the conditional expectations $\expar{ Y_T | W=1 }$ and $\expar{ Y_T | W=0 }$ can be efficiently computed via representation~\eqref{eq:exp_R1}. Here, the conditional distributions of the stopping times $[T|W=1]$ and $[T|W=0]$ are straightforward to obtain.

\subsection{Proof of Lemma~\ref{lem:min_max_cont_disc}} \label{subsec:proof_lem_min_max_cont_disc}

We first observe that, for every set $S \in {\cal K}$, its characteristic vector $\chi^S$ belongs to $\Psi$. Since $\bar{H}(r, \chi^S) = H(r,S)$, we infer that $\opt\eqref{eqn:prob_min_max_cont} \geq \opt\eqref{eqn:definition_U}$, and to conclude the desired claim, it suffices to prove the opposite inequality. To this end, we next observe that, for every $r \in \bbR$, its corresponding problem $\max_{\psi \in \Psi} \bar{H}(r, \psi)$ within $\eqref{eqn:prob_min_max_cont}$ is a linear optimization problem whose constraint matrix is totally unimodular, since $\Psi = \{ \psi \in [0,1]^n : \| \psi \|_1 = k \}$. Therefore, the latter problem has an integer optimal solution. Any such solution is a characteristic vector of some set in ${\cal K}$, meaning that $\max_{\psi \in \Psi} \bar{H}(r, \psi) = \max_{S \in {\cal K}} H(r,S)$, and in turn, $\opt\eqref{eqn:definition_U} \geq \opt\eqref{eqn:prob_min_max_cont}$.

\subsection{Proof of Lemma~\ref{lem:prob_Sm_Sp}}  \label{subsec:proof_lem_prob_Sm_Sp}

In what follows, we show that $\frac{ d }{ dr } H(r^*,S^-) \geq 0$. Nearly identical arguments can be employed to show that $\frac{ d }{ dr } H(r^*,S^+) \leq 0$, and we therefore omit these details. Suppose on the contrary that $\frac{ d }{ dr } H(r^*,S^-) < 0$. As such, by our definition of $S^-$, it follows that $\frac{ d }{ dr } H(r^*,S) \leq \frac{ d }{ dr } H(r^*,S^-) < 0$ for every subset $S \in {\cal F}^*$. In the remainder of this proof, we establish the next two claims:
\begin{enumerate}
    \item There exists $\eps_1 > 0$ such that $H(r,S) < \hmax( r^* )$ for all $S \in {\cal K} \setminus {\cal F}^*$ and $r \in [r^*,r^* + \eps_1]$.
    
    \item There exists $\eps_2 > 0$ such that $H(r,S) < \hmax( r^* )$ for all $S \in {\cal F}^*$ and $r \in (r^*,r^* + \eps_2]$.
\end{enumerate}
Given these results, we immediate infer that there exists $\hat{r} > r^*$ for which $\hmax( \hat{r} ) < \hmax( r^* )$, which is clearly impossible due to the optimality of $r^*$. For this purpose, letting $\hat{r} = r^* + \min \{ \eps_1, \eps_2 \}$, we have 
\begin{eqnarray*}
\hmax( \hat{r} ) & = & \max_{S \in {\cal K}} H( \hat{r}, S ) \\
& = & \max \left\{ \max_{S \in {\cal K} \setminus {\cal F}^*} H( r^* + \min \{ \eps_1, \eps_2 \}, S ), \max_{S \in {\cal F}^*} H( r^* + \min \{ \eps_1, \eps_2 \}, S ) \right\} \\
& < & \hmax( r^* ) \ ,
\end{eqnarray*}
where the last inequality follows from items~1 and~2 above. 

\paragraph{Proof of item~1.} Consider some subset $S \in {\cal K} \setminus {\cal F}^*$. By definition, we have $H(r^*,S) < \hmax( r^* )$. In addition, by Observation~\ref{obs:prop_H2_min}, we know that the function $H(\cdot,S)$ is convex, and hence, also continuous. Therefore, there exists $\eps_S > 0$ such that $H(r,S) < \hmax( r^* )$, for every $r \in [r^*,r^* + \eps_S]$. We conclude the proof by fixing $\eps_1 = \min_{ S \in {\cal K} \setminus {\cal F}^* } \eps_S$.

\paragraph{Proof of item~2.} Consider some subset $S \in {\cal F}^*$. By our initial assumption, $\frac{ d }{ dr } H(r^*,S) < 0$, meaning in particular that $\partial_+ H(r^*,S) = \lim_{\Delta \to 0^+} \frac{ H(r^* + \Delta, S) - H( r^*, S ) }{ \Delta } < 0$. Therefore, there exists $\Delta_S > 0$ such that $\frac{ H(r^* + \Delta, S) - H( r^*, S ) }{ \Delta } \leq \frac{ \partial_+ H(r^*,S)  }{ 2 }$ for all $\Delta \in (0,\Delta_S]$. By rearranging this inequality, we have
\[ H(r^* + \Delta, S) ~~\leq~~ H( r^*, S) + \frac{ \partial_+ H(r^*,S)  }{ 2 } \cdot \Delta ~~<~~ H( r^*, S) \ . \]
We conclude the proof by fixing $\eps_2 = \min_{ S \in {\cal F}^* } \Delta_S$.

\subsection{Proof of Lemma~\ref{lem:F_rstar_psistar}} \label{subsec:proof_lem_F_rstar_psistar}

We observe that, due to te linearity of $\bar{H}(r^*,\cdot)$,
\begin{eqnarray}
\bar{H}( r^*, \psi^*) & = & \bar{H}( r^*, \alpha \cdot \chi^{S^+} + (1 - \alpha) \cdot \chi^{S^-} ) \nonumber \\
& = & \alpha \cdot \bar{H}(r^*, \chi^{S^+} ) + (1 - \alpha) \cdot \bar{H}(r^*, \chi^{S^-} ) \nonumber \\
& = & \alpha \cdot H(r^*, S^+ ) + (1 - \alpha) \cdot H(r^*, S^- )  \nonumber \\
& = & \opt\eqref{eqn:definition_U} \label{eqn:proof_lem_F_rstar_psistar_eq1} \\
& = & \opt\eqref{eqn:prob_min_max_cont} \label{eqn:proof_lem_F_rstar_psistar_eq2} \\
& = & \max_{\psi \in \Psi} \bar{H}(r^*, \psi) \ . \label{eqn:proof_lem_F_rstar_psistar_eq3}
\end{eqnarray}
Here, equality~\eqref{eqn:proof_lem_F_rstar_psistar_eq1} holds since both $S^-$ and $S^+$ maximize $H(r^*,S)$ over all $S \in {\cal K}$, meaning that $H(r^*, S^+ ) = H(r^*, S^- ) = \opt\eqref{eqn:definition_U}$. Equality~\eqref{eqn:proof_lem_F_rstar_psistar_eq2} is precisely Lemma~\ref{lem:min_max_cont_disc}. Finally, Equality~\eqref{eqn:proof_lem_F_rstar_psistar_eq3} follows by recalling that $r^*$ is an optimal solution to problem~\eqref{eqn:prob_min_max_cont}.

\subsection{Proof of Lemma~\ref{lem:e_e-1_exp_B}} \label{subsec:proof_lem_e_e-1_exp_B}

In order to establish this equation, we observe that
\begin{eqnarray}
\ex{ B } & = & \sum_{i \in [k]} \ex{ B_i } \nonumber \\
& = & \sum_{i \in [k]} \pr{ Y_i \geq r^* } \nonumber \\
& = & \sum_{i \in [k] \setminus \{ \hat{i} \} } \pr{ X_{\pi(i)} \geq r^* } + \pr{ X_{\ell,m} \geq r^* } \nonumber \\
& = & \sum_{i \in [k] \setminus \{ \hat{i} \} } \pr{ X_{\pi(i)} \geq r^* } + \psi_{\ell}^* \cdot \pr{ X_{\ell} \geq r^* } + \psi_m^* \cdot \pr{ X_m \geq r^* } \label{eqn:proof_lem_e_e-1_exp_B_1} \\
& = & \sum_{i \in [n]} \pr{ X_i \geq r^* } \cdot \psi_i^* \nonumber \\
& = & 1 \ . \label{eqn:proof_lem_e_e-1_exp_B_2}
\end{eqnarray}
Here, equality~\eqref{eqn:proof_lem_e_e-1_exp_B_1} holds since $X_{\ell,m} = W X_{\ell} + (1-W) X_m$, where $W \sim \mybern( \psi^*_{ \ell })$ is independent of $X_{\ell}$ and $X_m$, noting that $\psi^*_{ \ell } + \psi^*_m = 1$, since these are the only two fractional coordinates of $\psi^*$. Equality~\eqref{eqn:proof_lem_e_e-1_exp_B_2} is precisely item~1 of Theorem~\ref{thm:structure_psi_star}.

\subsection{Proof of Lemma~\ref{lem:e_e-1_exp_min_binom}} \label{subsec:proof_lem_e_e-1_exp_min_binom}

To derive the desired inequality, note that
\begin{eqnarray}
\ex{ \min \{ B, 1 \} } & = & \pr{ B \geq 1 } \nonumber \\
& = & 1 - \prod_{i \in [k]} \left( 1 - \pr{ Y_i \geq r^* } \right) \label{eqn:proof_lem_e_e-1_exp_min_binom_1} \\
& = & 1 - \left( \prod_{i \in [k] \setminus \{ \hat{i} \} } \left( 1 - \pr{ X_{\pi(i)} \geq r^* } \right) \right) \cdot \left( 1 - \pr{ X_{\ell,m} \geq r^* } \right) \nonumber \\
& = & 1 - \left( \prod_{i \in [k] \setminus \{ \hat{i} \} } \left( 1 - \pr{ X_{\pi(i)} \geq r^* } \right) \right) \nonumber \\
&& \qquad \qquad \qquad \qquad \mbox{} \cdot \left( 1 - \psi_{\ell}^* \cdot \pr{ X_{\ell} \geq r^* } - \psi_m^* \cdot \pr{ X_m \geq r^* } \right) \label{eqn:proof_lem_e_e-1_exp_min_binom_2}  \\
& \geq & 1 - \exp \left\{ - \sum_{i \in [n]} \pr{ X_i \geq r^* } \cdot \psi_i^* \right\} \nonumber \\
& = & 1 - \frac{ 1 }{ e } \ . \label{eqn:proof_lem_e_e-1_exp_min_binom_3}
\end{eqnarray}
In this case, equality~\eqref{eqn:proof_lem_e_e-1_exp_min_binom_1} holds since $B_1, \ldots, B_k$ are mutually independent. Equality~\eqref{eqn:proof_lem_e_e-1_exp_min_binom_2} is obtained by recalling that $\prpar{ X_{\ell,m} \geq r^* } = \psi_{\ell}^* \cdot \prpar{ X_{\ell} \geq r^* } + \psi_m^* \cdot \prpar{ X_m \geq r^* }$, as explained when justifying equality~\eqref{eqn:proof_lem_e_e-1_exp_B_1}. Finally, equality~\eqref{eqn:proof_lem_e_e-1_exp_min_binom_3} follows from item~1 of Theorem~\ref{thm:structure_psi_star}, which shows that $\sum_{i \in [n]} \prpar{ X_i \geq r^* } \cdot \psi_i^* = 1$.

\subsection{Proof of Lemma~\ref{lem:e_e-1_exp_Rinfty}} \label{subsec:proof_lem_e_e-1_exp_Rinfty}

According to representation~\eqref{eq:exp_Rinfty_1} of the auxiliary random variable $Y^{ \Sigma }$, we have
\begin{eqnarray}
\ex{  Y^{ \Sigma } } & = & \sum_{i \in [k]} \pr{ B_i = 1 } \cdot \ex{ Y_i | Y_i \geq r^* } \nonumber \\
& = & \sum_{i \in [k]} \pr{ Y_i \geq r^* } \cdot \left( r^* + \ex{ Y_i - r^* | Y_i \geq r^* } \right) \nonumber \\
& = & r^* \cdot \sum_{i \in [k]} \pr{ Y_i \geq r^* } + \sum_{i \in [k] }  \pr{ X_{\pi(i)} \geq r^* } \cdot \ex{ X_{\pi(i)} - r^* | X_{\pi(i)} \geq r^* } \nonumber \\
& = & r^* + \sum_{i \in [k]} \ex{ [X_{\pi(i)} - r^*]^+ } \ , \label{eqn:proof_lem_e_e-1_exp_Rinfty_1}
\end{eqnarray}
where the last equality holds since $\sum_{i \in [k]} \pr{ Y_i \geq r^* } = \expar{B} = 1$, by Lemma~\ref{lem:e_e-1_exp_B}. Now, within the expression we have just obtained, note that $X_{\pi(\hat{i})} = X_{\ell,m} = W X_{\ell} + (1-W) X_m$, implying that  
\[ \ex{ [X_{\ell,m} - r^*]^+ } ~~=~~ \psi_{\ell}^* \cdot \ex{ [ X_{\ell} - r^* ]^+ } + \psi_m^* \cdot \ex{ [ X_m - r^* ]^+ } \ . \]
Therefore, equation~\eqref{eqn:proof_lem_e_e-1_exp_Rinfty_1} can be written as
\begin{eqnarray*}
\ex{  Y^{ \Sigma } } & = & r^* + \sum_{i \in [k] \setminus \{ \hat{i} \}} \ex{ [ X_{\pi(i)} - r^* ]^+ } + \psi_{\ell}^* \cdot \ex{ [ X_{\ell} - r^* ]^+ } + \psi_m^* \cdot \ex{ [ X_m - r^* ]^+ } \\
& = & r^* + \sum_{i \in [n] } G_i( r^* ) \cdot \psi_i^* \\
& = & \bar{H}( r^*, \psi^* ) \\
& = & \opt\eqref{eqn:prob_min_max_cont} \\
& = & \mybound \ .
\end{eqnarray*}
Here, the next-to-last equality holds since $r^*$ is an optimal solution to problem~\eqref{eqn:prob_min_max_cont}, whereas $\psi^*$ is optimal for its inner maximization problem, $\max_{\psi \in \Psi} \bar{H}(r^*, \psi)$, by Theorem~\ref{thm:structure_psi_star}. The last equality is obtained by recalling that $\mybound = \opt\eqref{eqn:definition_U} = \opt\eqref{eqn:prob_min_max_cont}$, according to Lemma~\ref{lem:min_max_cont_disc}.

\subsection{Proof of Lemma~\ref{lem:e_e-1_exp_diff}} \label{subsec:proof_lem_e_e-1_exp_diff}

By representation~\eqref{eq:exp_Rinfty_2}, we know that $Y^{ \Sigma } = Y_T + Y^{ \Sigma }_{ [T+1,k] }$, and therefore
\begin{eqnarray}
\ex{ Y^{ \Sigma } } - \ex{ Y_T } & = & \ex{ Y^{ \Sigma }_{ [T+1,k] } } \nonumber \\
& = & \sum_{t \in [k]} \pr{ T = t } \cdot \ex{ \left. Y^{ \Sigma }_{ [T+1,k] } \right| T = t } \nonumber \\
& = & \sum_{t \in [k]} \pr{ T = t } \cdot \sum_{i \in [t+1,k]} \ex{ B_i Y_i | T=t } \label{eqn:proof_lem_e_e-1_exp_diff_1} \\
& = & \sum_{t \in [k]} \pr{ T = t } \cdot \sum_{i \in [t+1,k]} \pr{ B_i = 1 | T=t } \cdot \ex{ Y_i | B_i = 1, T=t } \nonumber \\
& = & \sum_{t \in [k]} \pr{ T = t } \cdot \sum_{i \in [t+1,k]} \pr{ B_i = 1 } \cdot \ex{ Y_i | B_i = 1 } \label{eqn:proof_lem_e_e-1_exp_diff_2} \\
& = & \sum_{t \in [k]} \pr{ T = t } \cdot \underbrace{ \sum_{i \in [t+1,k]} \pr{ B_i = 1 } \cdot \ex{ Y_i | Y_i \geq r^* } }_{ (A_t) } \ . \nonumber
\end{eqnarray}
Here, equality~\eqref{eqn:proof_lem_e_e-1_exp_diff_1} follows by recalling that $Y^{ \Sigma }_{ [t+1,k] } = \sum_{i \in [t+1,k]} B_i Y_i$. Equality~\eqref{eqn:proof_lem_e_e-1_exp_diff_2} holds since $B_i$ is independent of the event $\{ T = t \} = \{ B_1 = \cdots = B_{t-1} = 0, B_t = 1\}$ when $i>t$; similarly, $[Y_i | B_i = 1]$ is independent of $\{ T=t \}$. In the next claim, whose proof is presented in Section~\ref{subsec:proof_clm_e_e-1_avg_arg}, we establish an upper bound on the term $(A_t)$, appearing in the latter expression.

\begin{claim} \label{clm:e_e-1_avg_arg}
$(A_t) \leq \mybound \cdot \sum_{i \in [t+1,k]} \pr{ B_i = 1 }$.
\end{claim}

Based on this result, we conclude the proof by observing that
\begin{eqnarray*}
\ex{ Y^{ \Sigma } } - \ex{ Y_T } & \leq & \mybound \cdot \sum_{t \in [k]} \pr{ T = t } \cdot \sum_{i \in [t+1,k]} \pr{ B_i = 1 } \\
& = & \mybound \cdot \sum_{t \in [k]} \pr{ T = t } \cdot \ex{ \left. [B-1]^+ \right| T = t } \\
& = & \mybound \cdot \ex{ [B-1]^+ } \ .
\end{eqnarray*}

\subsection{Proof of Claim~\ref{clm:e_e-1_avg_arg}} \label{subsec:proof_clm_e_e-1_avg_arg}

Our proof is based on an averaging argument, for which we define three independent random variables:
\begin{itemize}
    \item $Z$ takes the values $\expar{ Y_1 | Y_1 \geq r^* }, \ldots, \expar{ Y_k | Y_k \geq r^* }$ with probabilities $\prpar{ B_1 = 1 }, \ldots, \pr{ B_k = 1 }$, respectively. Noting that $\sum_{i \in [k]} \prpar{ B_i = 1 } = \ex{B} = 1$, by Lemma~\ref{lem:e_e-1_exp_B}, it follows that the distribution we have just defined is indeed valid. In addition,
    \begin{eqnarray}
    \ex{ Z } & = & \sum_{i \in [k]} \prpar{ B_i = 1 } \cdot \expar{ Y_i | Y_i \geq r^* } \nonumber \\
    & = & \sum_{i \in [k]} \prpar{ B_i = 1 } \cdot \expar{ Y_i | B_i = 1 } \nonumber \\
    & = & \sum_{i \in [k]} \ex{ B_i Y_i } \nonumber \\
    & = & \ex{ Y^{ \Sigma } } \nonumber \\
    & = & \mybound \ , \label{eqn:proof_clm_e_e-1_avg_arg_1}
    \end{eqnarray}
    where the last equality is exactly Lemma~\ref{lem:e_e-1_exp_Rinfty}.
    
    \item $Z^+$ takes the values $\expar{ Y_1 | Y_1 \geq r^* }, \ldots, \expar{ Y_t | Y_t \geq r^* }$ with probabilities $\prpar{ B_1 = 1 } / \sum_{i \in [t]} \prpar{ B_i = 1 }, \ldots, \pr{ B_t = 1 } / \sum_{i \in [t]} \prpar{ B_i  = 1 }$, respectively. When $\sum_{i \in [t]} \prpar{ B_i  = 1 } = 0$, the random variable $Z^+$ is defined as $\expar{ Y_1 | Y_1 \geq r^* }$ with probability $1$.

    \item $Z^-$ takes the values $\expar{ Y_{t+1} | Y_{t+1} \geq r^* }, \ldots, \expar{ Y_k | Y_k \geq r^* }$ with probabilities $\prpar{ B_{t+1} = 1 } / \sum_{i \in [t+1,k]} \prpar{ B_i = 1 }, \ldots, \pr{ B_k = 1 } / \sum_{i \in [t+1,k]} \prpar{ B_i  = 1 }$, respectively. Again, when $\sum_{i \in [t+1,k]} \prpar{ B_i  = 1 } = 0$, this random variable is defined as $0$ with probability $1$.
    \end{itemize}
By setting $\hat{W} \sim \mybern( \sum_{i \in [t]} \prpar{ B_i  = 1 })$, independently of $Z^+$ and $Z^-$, it is easy to verify that $Z$ and $\hat{W} Z^+ + (1 - \hat{W}) Z^-$ are identically distributed. In addition, since $\expar{ Y_1 | Y_1 \geq r^* } \geq \cdots \geq \ex{ Y_k | Y_k \geq r^* }$, we know that $Z^+ \geqst  Z^-$, implying in turn that $Z \geqst Z^-$. Given these observations, the desired upper bound on $(A_t)$ follows by noting that
\begin{eqnarray*}
(A_t) & = & \sum_{i \in [t+1,k]} \pr{ B_i = 1 } \cdot \ex{ Y_i | Y_i \geq r^* } \\
& = & \ex{Z^-} \cdot \sum_{i \in [t+1,k]} \pr{ B_i = 1 } \\
& \leq & \ex{Z} \cdot \sum_{i \in [t+1,k]} \pr{ B_i = 1 } \\
& = & \mybound \cdot \sum_{i \in [t+1,k]} \pr{ B_i = 1 } \ ,
\end{eqnarray*}
where the last equality holds since $\expar{Z} = \mybound$, as shown in~\eqref{eqn:proof_clm_e_e-1_avg_arg_1}.
\section{Concluding Remarks}

We conclude this paper by highlighting a number of open questions and potential directions for future research. The next few points are intended to investigate whether our adaptivity gaps can be sharpened within the adaptive ProbeMax problem by itself, as well as to examine whether our upper-bounding method for adaptive policies can be leveraged in broader settings.

\paragraph{General random variables: Constructive adaptivity gaps?} Prior to our work, the best-known constructive adaptivity gap for general random variables was $3$, via the Lagrangian relaxation approach of \cite{GuhaMS10}. While we have been successful at establishing an improved gap of $2$, including an explicit construction of a feasible set with matching expected maximum reward, it would be interesting to opt for additional improvements in this context. On possible direction could be reducing the general setting to the absolutely continuous case; for example, each of the random variables $X_i$ can be substituted by $\tilde{X}_i = X_i + \delta_i$, where say $\delta_i \sim U(0,\Delta)$ with $\Delta \ll \mu_{\max}$, as suggested in Section~\ref{subsec:prev_work_open_questions}. However, from an algorithmic standpoint, our approach requires oracle access to $\prpar{ \tilde{X}_i \leq \cdot}$ and $\expar{ \tilde{X}_i | \tilde{X}_i \geq \cdot }$. While we can design several forms of approximate oracles through their exact counterparts with respect to the original random variables $X_i$, it is unclear whether the algorithmic ideas of Section~\ref{sec:adapt_gap_e_em1} and their analysis are robust to such approximation errors.

\paragraph{Extensions to additional feasibility constraints?} Taking a broader perspective, the adaptive ProbeMax problem falls within the framework of stochastic probing subject to downward-closed constraints, capturing much of its analytical and computational challenges. As reported in Section~\ref{subsec:prev_work_open_questions}, \cite{GuptaN13} proposed an LP-based randomized rounding approach, creating a distribution over feasible sets whose expected objective value approximates the adaptive optimum within factor $3$. These results are applicable in the discrete case, with finite support, assuming that the specific constraints in question allow us to efficiently solve this LP-relaxation, which is indeed doable for matroid, knapsack, and $k$-system constraints, to mention a few. As part of future research, it would be interesting to study whether suitable adaptations of our min-max upper bound can be carried over to the downward-closed setting. Particularly relevant questions in this context are those of obtaining improved adaptivity gaps, deterministic constructions of non-adaptive policies, and simple threshold-based policies.

\paragraph{Concurrent work.} Prior to the journal submission of this paper, we have learned about the work of \cite{EpsteinM22}, who have independently considered the adaptive Probe-$\ell$-Max problem. This setting has the objective of maximizing the expected sum of $\ell$-largest rewards rather than only the maximal one. While both papers aim at deriving constructive adaptivity gaps, they are very different in terms of their scope and methodology. Specifically, \citeauthor{EpsteinM22} focus on the case where $X_1, \ldots, X_n$ are discrete random variables with finite support, showing that
optimal fractional solutions to the linear program described in Section~\ref{subsec:prev_work_open_questions} can be rounded to a non-adaptive probing policy whose expected total reward is within factor $1 - e^{-\ell} \cdot \frac{ \ell^{\ell} }{ \ell! }$ of the adaptive optimum. While this result is incomparable with our adaptivity gaps for general random variables as well as for continuous ones, it improves on the currently best known gaps in the discrete setting. For a thorough comparison between their findings and earlier results along these lines \citep{GuhaM07, GuptaN13, GuptaNS17, Bradac0Z19}, we refer the avid reader to Section~1 in~\citep{EpsteinM22}.

\paragraph{Acknowledgements.} We are grateful to Viswanath Nagarajan (University of Michigan) and Will Ma (Columbia University) for several technical discussions and for additional pointers to earlier literature.

\addcontentsline{toc}{section}{Bibliography}
\bibliographystyle{plainnat}
\bibliography{BIB-Adaptive-Probemax}

\end{document}